\documentclass[preprint]{aastex62}
\usepackage{amsmath,amstext}
\usepackage{graphicx}

\def\vec#1{\ensuremath{\mathbf{#1}}}

\newcounter{RomanNumber}
\newcommand{\MyRoman}[1]{\setcounter{RomanNumber}{#1}\Roman{RomanNumber}}

\received{****}
\revised{****}
\accepted{****}

\shorttitle{Fe XXI 1354 \AA\ Emissions Modulated by Sausage Modes}
\shortauthors{Shi et al.}
\begin{document}

\title{Synthetic Emissions of the Fe XXI 1354 \AA\ Line from Flare Loops Experiencing Fundamental Fast Sausage Oscillations}

\correspondingauthor{Bo Li}
\email{bbl@sdu.edu.cn}

\author{Mijie Shi}
\affiliation{Shandong Provincial Key Laboratory of Optical Astronomy and Solar-Terrestrial Environment, Institute of Space Sciences, Shandong University, Weihai 264209, China}
\affiliation{CAS Key Laboratory of Solar Activity, National Astronomical Observatories, Beijing 100012, China}

\author{Bo Li}
\affiliation{Shandong Provincial Key Laboratory of Optical Astronomy and Solar-Terrestrial Environment, Institute of Space Sciences, Shandong University, Weihai 264209, China}

\author{Zhenghua Huang}
\affiliation{Shandong Provincial Key Laboratory of Optical Astronomy and Solar-Terrestrial Environment, Institute of Space Sciences, Shandong University, Weihai 264209, China}


\author{Shao-Xia Chen}
\affiliation{Shandong Provincial Key Laboratory of Optical Astronomy and Solar-Terrestrial Environment, Institute of Space Sciences, Shandong University, Weihai 264209, China}



\begin{abstract}
Inspired by recent IRIS observations,
   we forward model the response of the Fe XXI 1354 \AA\ line
   to fundamental, standing, linear fast sausage modes (FSMs) in flare loops. 
Starting with the fluid parameters for an FSM in a straight tube with equilibrium parameters largely compatible with
   the IRIS measurements, we synthesize the line profiles
   by incorporating the non-Equilibrium Ionization (NEI) effect in the computation of the contribution function. 
We find that both the intensity and Doppler shift oscillate at the wave period ($P$).
The phase difference between the two differs from the expected value ($90^\circ$)
   only slightly because 
   {NEI plays only a marginal role in determining the ionic fraction of Fe XXI
   in the examined dense loop}.
The Doppler width modulations, however, posses an asymmetry in the first and second halves of a wave period,
   leading to a secondary periodicity at $P/2$ in addition to the 
   primary one at $P$.
This behavior results from 
   the competition between the broadening due to bulk flow and that due to temperature variations,
   with the latter being stronger but not overwhelmingly so.
These expected signatures, with the exception of the Doppler width, 
   are largely consistent with the IRIS measurements, thereby corroborating the reported 
   detection of a fundamental FSM.
The forward modeled signatures are useful for identifying fundamental FSMs in flare loops
   from measurements of the Fe \MyRoman{21} 1354 \AA\ line with instruments similar to IRIS,
   even though a much higher cadence is required for the expected behavior in the Doppler widths
   to be detected. 
\end{abstract}

\keywords{magnetohydrodynamics --- Sun: corona --- Sun: UV radiation --- waves}



\section{Introduction} 
\label{sec:intro}
Magnetohydrodynamic (MHD) waves and oscillations have been amply found 
   in the highly structured solar corona
   \citep[see recent reviews by e.g.,][]
	{2005LRSP....2....3N, 2007SoPh..246....3B, 2012RSPTA.370.3193D, 2016SSRv..200...75N}.
Theoretically, fast sausage modes (FSMs) are characterized by the absence of azimuthal dependence
   of the associated perturbations 
   and also by their axial phase speeds being on the order of the Alfv\'en ones
	\citep[e.g.,][]{1983SoPh...88..179E}.
FSMs can be either leaky or trapped depending on the longitudinal wavenumber $k$
   (\citeauthor{1986SoPh..103..277C}~\citeyear{1986SoPh..103..277C},
   also see e.g., 
   \citeauthor{1970A&A.....9..159R}~\citeyear{1970A&A.....9..159R};
   \citeauthor{1975IGAFS..37....3Z}~\citeyear{1975IGAFS..37....3Z};
   \citeauthor{1978SoPh...58..165M}~\citeyear{1978SoPh...58..165M};
   \citeauthor{2005A&A...441..371T}~\citeyear{2005A&A...441..371T};
   and \citeauthor{2007AstL...33..706K}~\citeyear{2007AstL...33..706K}).
Considerable efforts have gone into examining 
   how FSMs in coronal structures depend on the spatial distribution of the equilibrium parameters,
   thereby addressing such effects as the density distribution
   \citep[e.g.,][]{2009A&A...503..569I, 2012ApJ...761..134N, 2015ApJ...812...22C, 2015ApJ...814...60Y, 
   2016ApJ...833..114C, 2018JPhA...51b5501C},
   siphon flow~\citep[e.g.,][]{2003SoPh..217..199T, 2013ApJ...767..169L, 2014A&A...568A..31L},
   and magnetic twist~\cite[e.g.,][]{1999SoPh..185...41B, 2007SoPh..246..101E, 2012SoPh..280..153K}.
Observationally, FSMs are often invoked to interpret quasi-periodic pulsations (QPPs) with quasi-periods
   ranging from a few seconds to a couple of tens of seconds in solar flares 
   \citep[for recent reviews, see e.g.,][]{2009SSRv..149..119N,2016SoPh..291.3143V, 2018SSRv..214...45M}.
For instance, the existence of fundamental standing FSMs in flare loops was inferred from 
   the radio pulsations for a substantial number of flares measured by 
   the Nobeyama Radioheliograph~\citep[e.g.,][]{2003A&A...412L...7N, 2005A&A...439..727M, 2015A&A...574A..53K}.
Likewise, signatures of FSMs were also found in the fine structures 
   in a number of flare-associated radio bursts as measured by, say,
   the Chinese Solar Broadband Radio Spectrometer~\citep{2013ApJ...777..159Y, 2016ApJ...826...78Y},
   the Assembly of Metric-band Aperture Telescope and Real-time Analysis System~\citep{2018ApJ...855L..29K},
   and the LOw Frequency ARray~\citep{2018ApJ...861...33K}.
Furthermore, a recent multi-wavelength study by \citet{2018ApJ...859..154N} suggested the existence 
   of FSMs in loops associated with microflares.  
In extreme ultraviolet (EUV), \citet{2012ApJ...755..113S} 
   identified FSMs using imaging observations in  the 171 \AA\ channel
   of the Atmospheric Imaging Assembly (AIA) aboard the Solar Dynamics Observatory (SDO).
Recently, the oscillatory behavior in the profile of the Fe XXI 1354 \AA\ line
   as measured by the Interface Region Imaging Spectrograph (IRIS)
   was attributed to the fundamental standing FSM hosted by flare loops
   \citep[][hereafter T16]{2016ApJ...823L..16T}.

Forward modeling plays an important role in the practice of wave mode identification,
   which is often not straightforward in the case of FSMs.
\citet{2003A&A...397..765C} and \citet{2012A&A...543A..12G} 
   integrated the squared density along a line-of-sight (LoS) to compute
   the intensity modulations, thereby examining the effects of, say, 
   instrumental resolutions on the detectability of FSMs in coronal loops.
These effects on the synthetic profiles of the Fe IX 171~\AA\
   and Fe XII 193~\AA\ lines were 
   then examined, in a rather exhaustive manner,
   by \citet{2013A&A...555A..74A} who incorporated the relevant contribution functions.
As found from this series of studies,
   the modulations in the intensities, Doppler shifts, and Doppler widths
   are detectable even when the pixel size (temporal cadence) is 
   a substantial fraction of the axial wavelength (wave period).
Still addressing the Fe IX 171~\AA\
   and Fe XII 193~\AA\ lines, our recent study further examined
   how non-equilibrium ionization (NEI) affects the spectral signatures of FSMs in active region loops~\citep[][hereafter paper I]{2019ApJ...870...99S}. 
It was found that NEI has little effect on the Doppler shift or width,
   but tends to substantially reduce the intensity modulations, making the detection of FSMs in tenuous loops 
   possibly challenging.

This study aims to extend our paper I by forward modeling the spectral signatures 
   of fundamental FSMs in flare loops for the Fe {\rm XXI} 1354~\AA\ line, 
   as inspired by T16.
Such a study has not appeared in the literature to our knowledge, and is expected to find applications
   for identifying FSMs with the rich information that this flare line can offer. 
This manuscript is organized as follows.
In Section~\ref{sec_model_formulation} we present the necessary 
   expressions for the perturbations associated with linear FSMs,
   and describe the procedure for synthesizing 
   the Fe {\rm XXI} 1354~\AA\ line profiles.
The synthetic emission properties are then detailed in 
   Section~\ref{Results}.
Section~\ref{summary} summarizes this study, ending with some concluding remarks.

\section{Model Formulation}
\label{sec_model_formulation}

\subsection{Fundamental Fast Sausage Modes In Flare Loops}
\label{sec_sausage_model}
We model an equilibrium flare loop as a static, straight cylinder
    with mean radius $R=5\times10^3$~km and length $L = 4.5 \times 10^4$~km. 
In a cylindrical coordinate system ($r, \phi, z$),
    both the cylinder axis and the equilibrium magnetic field ${\mathbf B}$
    are in the $z$-direction.
We adopt single-fluid ideal MHD and consider an electron-proton plasma throughout.
Gravity is neglected.
We further assume that the equilibrium parameters
    depend only on $r$, and in such a way that the configuration comprises
    a uniform cord (denoted by subscript $\rm i$),
    a uniform external medium (subscript $\rm e$),
    and a transition layer (TL) continuously connecting the two.
Let $N$, $T$, and $B$ denote the electron number density,
    electron temperature, and magnetic field strength, respectively.
Furthermore, let subscript $0$ denote the equilibrium values.    
In the cord, we take 
    $N_{\rm i} = 5 \times 10^{10} {\rm~cm}^{-3}$,
    $T_{\rm i} = 10~{\rm MK}$,
    and $B_{\rm i} = 43~{\rm G}$,
    resulting in an internal Alfv\'en speed $v_{\rm Ai}$ of $420~{\rm km}~{\rm s}^{-1}$. 
We assume that $N_0$ and $T_0$ in the TL depend linearly on $r$,
    their external values being $1.1 \times 10^{9} {\rm~cm}^{-3}$
    and $2~{\rm MK}$, respectively.
This TL is of width $R$ and centered around $r=R$.    
The $r$-dependence of $B_0$ then follows from the transverse force balance,
    yielding a $B_{\rm e}$ of $72.8~{\rm G}$.

Consider standing linear FSMs in this equilibrium,
    and suppose that the system has reached a stationary state
    characterized by angular frequency $\omega$
    and axial wavenumber $k$.
The physical variables relevant for computing UV emissions are given by
\begin{eqnarray}
&& N (r, z; t)    =
    N_0 \left\{1-
            \left[\frac{d \ln N_0(r)}{d r} {\cal R}(r) + {\cal D}(r)
            \right] \sin(\omega t)\sin(kz)
        \right\}~,
         \label{eq_pert_den} \\[0.1cm]
&& v_r (r, z; t)  =
        \omega {\cal R}(r) \cos(\omega t)\sin(kz)~,
         \label{eq_pert_vr} \\[0.1cm]
&& \displaystyle 
   v_z (r, z; t)  =
      -\frac{c_{\rm s}^2}{\omega/k} {\cal D}(r) \cos(\omega t)\cos(kz)~,
         \label{eq_pert_vz} \\[0.1cm]
&& T (r, z; t)  =
      T_0 \left\{1-
            \left[\frac{d \ln T_0(r)}{d r} {\cal R}(r) + (\gamma-1){\cal D}(r)
            \right] \sin(\omega t)\sin(kz)
        \right\}~,
         \label{eq_pert_T}
\end{eqnarray}    
    where $\gamma = 5/3$ is the adiabatic index, and $c_{\rm s} = \sqrt{2 k_{\rm B} T_0/m_{\rm p}}$
    gives the adiabatic sound speed with $k_{\rm B}$ being the Boltzmann constant and $m_{\rm p}$
    the proton mass. 
Furthermore, ${\cal R}$ is connected to the transverse (i.e., radial) Lagrangian displacement,
    and ${\cal D}$ is related to ${\cal R}$ by 
\begin{eqnarray}
 {\cal D} (r) = \frac{\omega^2}{\omega^2-k^2 c_{\rm s}^2}
                \frac{{\rm d} (r {\cal R})}{r {\rm d} r}~.
\label{eq_compressibity}
\end{eqnarray}
Given the equilibrium configuration, a single equation governing ${\cal R}$
    can be readily derived from the linearized ideal MHD equations
    \citep[see Equation~15 in][hereafter C16]{2016ApJ...833..114C}.
This equation is then solved separately in the internal and external media
    as well as in the TL, enabling the derivation of a dispersion relation (DR)
    in view of the continuity of both ${\cal R}$ and its derivative
    at the relevant interfaces. 
Once a longitudinal wavenumber $k$ is supplied,     
    solving this DR (Equation~23 in C16) then yields     
    the angular frequency $\omega$, which can then be employed to determine ${\cal R}$
    to within a constant factor characterizing the wave amplitude. 
For completeness, the Lagrangian displacements in the radial
    and axial directions are given by
\begin{eqnarray}
&& \xi_r (r, z ; t) = 
        {\cal R}(r) \sin(\omega t) \sin(kz)~, 
         \label{eq_xir} \\ 
&& \displaystyle 
   \xi_z (r, z ; t) = 
         -\frac{c_{\rm s}^2}{\omega^2/k} {\cal D}(r) \sin(\omega t)\cos(kz)~.
         \label{eq_xiz}
\end{eqnarray}

At this point, we note that the values for the equilibrium parameters are 
    largely compatible with the IRIS loop examined by T16.    
On the other hand, their $r$-dependence in the TL and the TL width 
    are admittedly difficult to constrain observationally.
Then why do we complicate the problem by adopting a continuous radial profile for the equilibrium parameters rather than 
    a piece-wise constant one as in \citet{2013A&A...555A..74A} and \citet{2019ApJ...870...99S}?
The reason is, if one would like to synthesize the specific intensities of 
    optically thin emissions, as we shall do, piece-wise constant profiles may yield inaccurate values.
This point will be detailed elsewhere, but the basic idea can be readily illustrated by considering the simplest case where
    the synthetic intensity $I$ is approximated by~\citep[e.g.,][]{2012A&A...543A..12G} 
\begin{eqnarray}
I(t) \propto \int_{\rm LoS} N^2 (\vec{x}-\vec{\xi}, t) dl = \int_{\rm LoS} (N_0^2 + 2 N_0 \Delta N + \Delta N^2)(\vec{x}-\vec{\xi}, t) dl ~.
\label{eq_I_densqrd}
\end{eqnarray}
Here $\vec{x}$ denotes a (fixed) location along an LoS, and $\vec{\xi}(\vec{x}, t)$ is the displacement vector.
Evaluating $N$ (and other flow parameters) this way makes sure that $N$ attains the value at a location $\vec{x}_P$
    that is displaced onto the LoS by $\vec{\xi}$.
\footnote{Ideally $\vec{x}_P$ should be computed such that $\vec{x}_P + \vec{\xi} (\vec{x}_P, t) = \vec{x}$.
However, this equation is implicit in $\vec{x}_P$ and therefore needs to be solved by iteration.
We refrain from doing so because it is time-consuming, and more importantly, the value of $\vec{x}_P$
    thus found differs from $\vec{x} - \vec{\xi}(\vec{x}, t)$ by only a second-order term.
}
In addition, we have decomposed the electron density $N$ into the equilibrium value $N_0$ 
    and the perturbation $\Delta N$.
Without loss of generality, 
    let us consider an LoS that is perpendicular to the cylinder axis.
Suppose further that the spatial scale of the LoS projected on to the plane-of-sky is much smaller than the axial wavelength,
    meaning that the leading-order perturbation in the integrand in Equation~\eqref{eq_I_densqrd} survives the integration process.
One now readily recognizes from Equation~\eqref{eq_pert_den}
    that $\Delta N$ involves two terms, one ($\Delta N_1$) proportional to ${\cal R} d N_0/dr$ and the other ($\Delta N_2$) 
    to $N_0 {\cal D}$.   
Note that while both terms need to be retained to maintain mass conservation, $\Delta N_1$ is 
    nominally discarded for synthesizing emissions modulated by FSMs in coronal tubes with
    piece-wise constant profiles ($d N_0/dr = 0$ in both the interior and exterior).
The LoS integration of $2 N_0 \Delta N_1 = {\cal R}dN_0^2/dr$, however, does not vanish because 
    $\cal {R}$ is finite and $dN_0^2/dr$ diverges at the interior-exterior interface. 
We note that this finite contribution to the intensity modulation can be incorporated analytically
    if one computes $I$ with Equation~\eqref{eq_I_densqrd}. 
However, strictly speaking, Equation~\eqref{eq_I_densqrd} holds only if
    the contribution function ($G_{\lambda 0}$) is position-independent
    (see Equation~\ref{eq_def_eps}).
In reality, $G_{\lambda 0}$ may be spatially dependent.
When equilibrium ionization (EI) is assumed, this dependence is primarily through the sensitive dependence of 
    $G_{\lambda 0}$ on electron temperature ($T$).
When the effect of non-equilibrium ionization (NEI) is substantial, the spatial dependence of $G_{\lambda 0}$ can also come from
    its dependence on electron density ($N$). 
In either case, $G_{\lambda 0}$ does not depend on $T$ and/or $N$ in a manner simple enough for one
    to readily incorporate the contribution to $I$ from $\Delta N_1$.
This makes it difficult to quantify the importance of this contribution relative to the one due to $\Delta N_2$. 
In contrast, adopting a continuous profile for the equilibrium parameters in the first place is 
    more self-consistent from the theoretical viewpoint, 
    and much easier to implement than correcting for the $\Delta N_1$ term 
    afterwards in practice. 
 
The following parameters are adopted for the perturbations.
The axial wavenumber $k$ is taken to be $\pi/L$, corresponding to
    the longitudinal fundamental mode as inspired by the IRIS observation.
We consider only the transverse fundamental, namely, the one for which $\cal R$ possesses 
    the simplest radial dependence. 
This mode is in the trapped regime, and corresponds to  
    a wave period $P = 2\pi/\omega$ of $19.1$~secs, which is close to the period
    in the Fe XXI 1354 \AA\ oscillations 
    examined by T16.
Consequently, the axial phase speed $\omega/k = 2 L/P$ reads 
    $4710$~km~s$^{-1}$ or $\sim 11.2~v_{\rm Ai}$.
The wave amplitude is specified such that the radial speed ($v_r$) attains a maximum of 
    $10.5$~km~s$^{-1}$, or equivalently $\sim 0.025 v_{\rm Ai}$.
As a result, the peak value in the perturbed density (temperature)
    reads $\sim 0.027 N_{\rm i}$ ($\sim 0.018 T_{\rm i}$).
The axial speed ($v_z$) reaches up to $v_z = 1.6$~km~s$^{-1}$.
The reason for us to choose such a small wave amplitude is that 
    the Fe XXI 1354~\AA\ intensity varies by only a few percent
    as measured by T16 (see Figure~3B therein).

\begin{figure}
  \begin{centering}
  \includegraphics[width=0.8\linewidth]{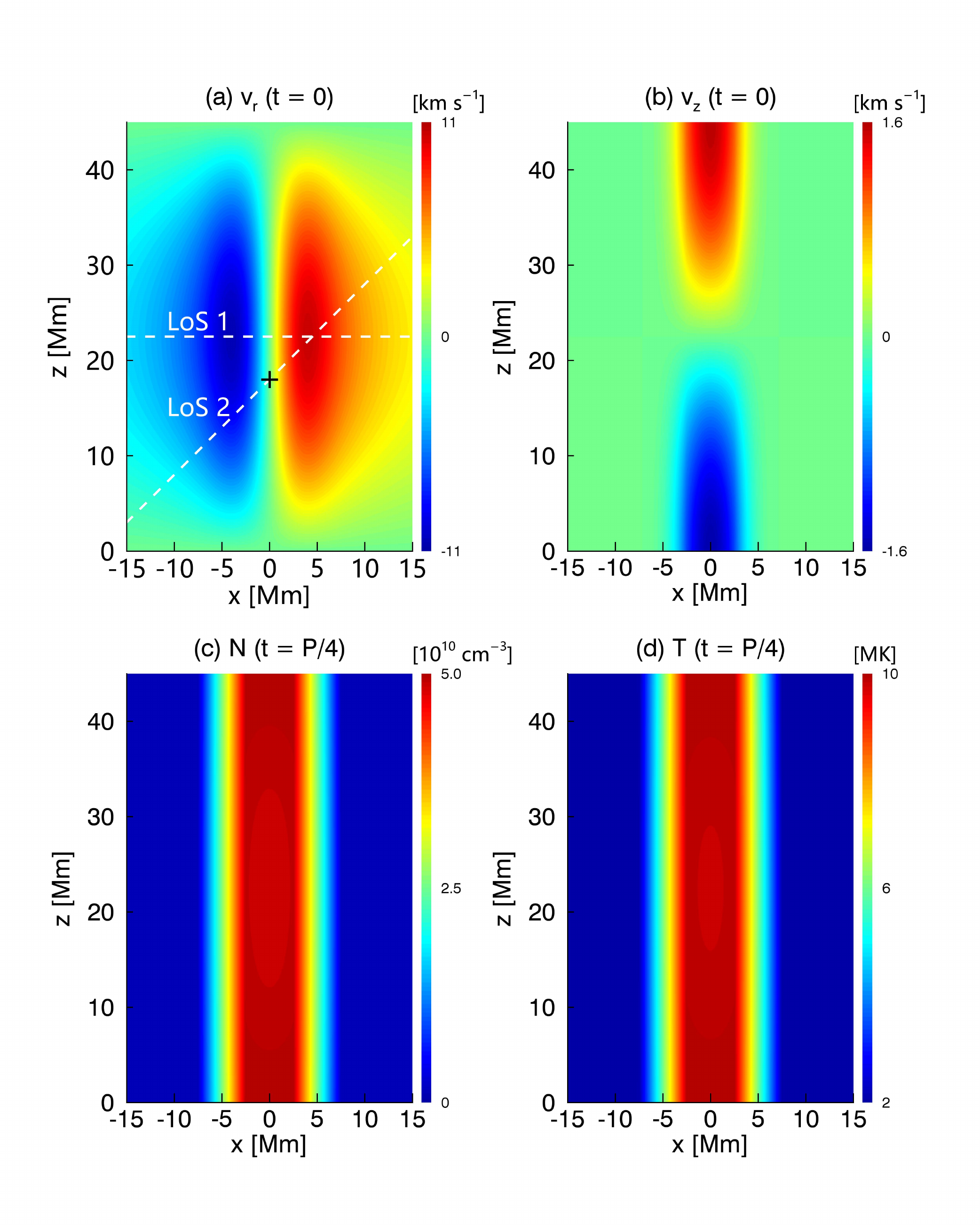}
  \caption{Snapshots of spatial distributions
    of the fluid parameters associated with the fast sausage mode.
  Shown here is a cut through the cylinder axis.
  The radial ($v_r$, panel a) and axial ($v_z$, panel b) velocities are for $t=0$,
      while the electron number density ($N$, panel c) and temperature ($T$, panel d)
        are for $t$ being one quarter of the wave period.
  All lines of sight adopted in this study are parallel to the $x-z$ plane,
      pertaining to slits along the $y$-direction.
  Two representative lines of sight, both in the $x-z$ plane, 
      are shown by the white dashed lines.
  LoS 1 is perpendicular to the loop axis and passes through the loop apex,
      while LoS 2 makes an angle of $45^\circ$ with the loop 
      and intersects the loop axis at $z_0 = 0.4 L$.
  }
  \label{f1}
 \end{centering}
\end{figure}

The spatial distributions of
    the fluid parameters in the $r-z$ plane are then constructed 
    with a spacing of $20$~km
    in both directions for $t$ between $0$ and $4$ periods.
Advection is accounted for by employing Equations~\eqref{eq_xir} and \eqref{eq_xiz}
    during this implementation, even though 
    a Eulerian grid is chosen for convenience (see paper I for details).
Figure \ref{f1} then presents these distributions in a cut 
    through the cylinder axis, with $x$ denoting an arbitrarily chosen perpendicular direction.
The radial ($v_r$) and axial velocities ($v_z$) are shown for $t=0$, 
    while the electron number density ($N$) and temperature ($T$) are shown for $t=P/4$.
Different instants of time are necessary because of the $\pi/2$ phase difference between
    the relevant perturbations (see Equations~\ref{eq_pert_den}
    to \ref{eq_pert_T}).
The continuous distribution of the equilibrium density or temperature can be readily seen,
    which, however, makes the expansion of the cylinder hard to tell
    (see Figure~\ref{f1}c or \ref{f1}d).
This happens because the maximal radial displacement is merely $\sim 32$~km.    

Without loss of generality, we assume that all lines of sight are parallel 
    to the $x-z$ plane, with two representative ones shown by the white dashed lines 
    in Figure~\ref{f1}.
When placed in the context of spectroscopic measurements, this means that we are placing slits
    along the $y$-direction. 
It then follows that an LoS is describable by three geometric parameters, one
    being the viewing angle $\theta$ between the LoS and the cylinder axis, 
    and the other two being the combination $[y_0, z_0]$ that characterizes where
    the LoS intersects the $y-z$ plane. 
For instance, LoS 1 pertains to a $\theta$ being $90^\circ$ 
    and $[y_0, z_0] = [0, 0.5 L]$.
Another one (LoS 2) corresponds to a $\theta$ of $45^\circ$ and 
    $[y_0, z_0] = [0, 0.4 L]$.

\subsection{Forward Modeling}
\label{sec_forward_model}

The emissivity of the Fe \MyRoman{21} 1354 \AA\ line is calculated at each grid point in the $r-z$ plane via 
\begin{eqnarray}
\label{eq_def_eps}
   \epsilon = G_{\lambda0} N^2~,
\end{eqnarray}
   where the contribution function $G_{\lambda0}$ is given by
\begin{eqnarray}
\label{eq_def_G}
G_{\lambda0} = h\nu_{ij}\cdot0.83\cdot Ab({\rm{Fe}}) f_{\mathrm{XXI}} \frac{n_jA_{ji}}{N}~.
\end{eqnarray}
Here $h\nu_{ij}$ is the energy level difference,
	$Ab({\rm{Fe}})$ is the abundance of Fe relative to Hydrogen,
	$f_{\mathrm{XXI}}$ is the ionic fraction of Fe \MyRoman{21},
	$n_j$ is the fraction of Fe \MyRoman{21} lying in the excited state,
	and $A_{ji}$ is the spontaneous transition probability.
We compute $G_{\lambda 0}$ using the function \textbf{g\_of\_t} from the CHIANTI package,
    with the ionic fraction $f_{\mathrm{XXI}}$ obtained by solving 
\begin{eqnarray}
 \label{eq_ionic_frac}
 \displaystyle
 \left(\frac{\partial}{\partial t}+{\mathbf v}\cdot\nabla\right) f_q
 = N \left[f_{q-1}C_{q-1}-f_q\left(C_q+R_q\right)+f_{q+1}R_{q+1}\right]~,
\end{eqnarray}
    where the ionization ($C$) and recombination ($R$) rate coefficients are found from CHIANTI as well
   (ver~8, \citeauthor{2015A&A...582A..56D}~\citeyear{2015A&A...582A..56D})
   \footnote{http://www.chiantidatabase.org/}.

Two remarks are necessary here.
First, lying behind the procedure outlined above for computing the emissivities is the coronal model approximation.
In particular, the ground level of an ion is assumed to be by far the most populated,
    which may not be true for Fe XXI when the electron density ($N$) is sufficiently high such that populations of metastable levels
    are no longer negligible.
This may indeed be a concern because we are examining the forbidden Fe XXI 1354~\AA\ line.
Fortunately, the outlined procedure still applies because the values adopted for $N$ 
    are still well below the critical value of $10^{12}~{\rm cm}^{-3}$~\citep[][Table 4]{2006ApJS..162..261L}.
Second, Equation~\eqref{eq_ionic_frac} reduces to the much simpler Equilibrium-Ionization (EI) case   
    when the wave period is much longer 
    than the ionization and recombination timescales.
We find that the Non-EI effect is marginal for Fe XXI for the examined dense flare loop,
    in contrast to the cases of Fe IX and Fe XII pertaining to typical active region loops as examined in paper I.
In practice, we nonetheless solve Equation~\eqref{eq_ionic_frac} for safety
    {and assess the role that NEI plays afterwards}.
{The solution procedure is initiated with the ionic fractions pertinent to the EI case 
	as in paper I~\citep[see also][for similar computations but in other contexts]{2010ApJ...722..625K, 2013ApJ...773..110S}.
It then follows that the computed ionic fractions, and hence the spectral parameters, 
    will show no difference from the EI case if ionization equilibrium can indeed be maintained 
    at all times.
}

Our forward modeling study then proceeds
   by converting the computed data ($\epsilon$, $T$, $v_r$, and $v_z$) from the cylindrical to 
   the aforementioned Cartesian coordinate system, covering
   a grid with spacing of $20$~km in all three directions.
To find the spectral profiles of the Fe \MyRoman{21} 1354 \AA\ line,   
   we then evaluate, at each grid point, the monochromatic emissivity $\epsilon_{\lambda}$
   at wavelength $\lambda$ as given by~\citep[see e.g.,][]{2016FrASS...3....4V}
\begin{eqnarray}
   \epsilon_\lambda = \frac{2\sqrt{2\ln2}}{\sqrt{2\pi}\lambda_w}\epsilon
      \exp\left\{
         -\frac{4\ln2}{\lambda_w^2}\left[
                                   \lambda-\lambda_0\left(1-\frac{v_{\rm LoS}}{c}\right)
                                   \right]^2
         \right\}~.
\label{eq_def_eps_lambda}
\end{eqnarray}
Here $\lambda_w=(2\sqrt{2\ln2})\lambda_0 (v_{\rm th}/c)$ is the thermal width,
   and $v_{\rm th}$ is the thermal speed determined by the instantaneous temperature. 
Furthermore, $\lambda_0$ is the rest wavelength, and    
   $v_{\rm LoS}$ the instantaneous velocity projected onto an LoS.
In what follows, by ``intensity'' and ``monochromatic intensity'' we mean 
\begin{eqnarray}
   I = \int_{\rm LoS} \frac{\epsilon}{4\pi} dl~,
   \hskip 0.5cm 
   \mbox{and}
   \hskip 0.5cm 
   I_{\lambda} = \int_{\rm LoS} \frac{\epsilon_{\lambda}}{4\pi} dl~,
\label{eq_def_I_I_lambda}   
\end{eqnarray}
   respectively.
Evidently, $\int I_{\lambda} d\lambda = I$.
By ``spectral profile'', we mean the dependence of $I_\lambda$ on $\lambda$.
With IRIS in mind, we assume that when projected onto the plane-of-sky, 
   any LoS corresponds to a square of $0.33\arcsec$ by $0.33\arcsec$
   (or $240$~km $\times$ $240$~km). 
In addition, $I_\lambda$ is sampled at a spacing of $26$~m\AA\
   for $\lambda$ between $\lambda_0-0.78$~\AA\ and $\lambda_0$+0.78~\AA.	
When performing the LoS integration for either $I$ or $I_\lambda$,
   we discretize an LoS into a series of thin beams, equally spaced by $20$~km
   and each with a size of $20$~km by $20$~km when projected onto the plane-of-sky.
For each thin beam, the integration is conducted on a grid of points equally spaced by $20$~km, 
   with the points labeled by their LoS coordinate $l$
   ($l=0$ pertains to where an LoS intersects the $y-z$ plane,
   see the cross in Figure~\ref{f1}).
The values thus found are then averaged over all thin beams, yield $I$ or $I_\lambda$.
From the derived $I_{\lambda}$, we then derive the Doppler velocity $v_{\rm D}$
    and width $w_{\rm D}$ by conducting a standard Gaussian fitting.
When constructing the distribution of the spectral parameters 
   along a slit, we employ a series of parallel lines-of-sight
   equally spaced by $240$~km.
This yields $I$, $v_{\rm D}$ and $w_{\rm D}$ as functions of $y$, which labels
   the locations along a slit (see Figures~\ref{f1} and \ref{f2}).

\section{Results}
\label{Results}
\begin{figure}
\begin{centering}
 \includegraphics[width=0.8\linewidth]{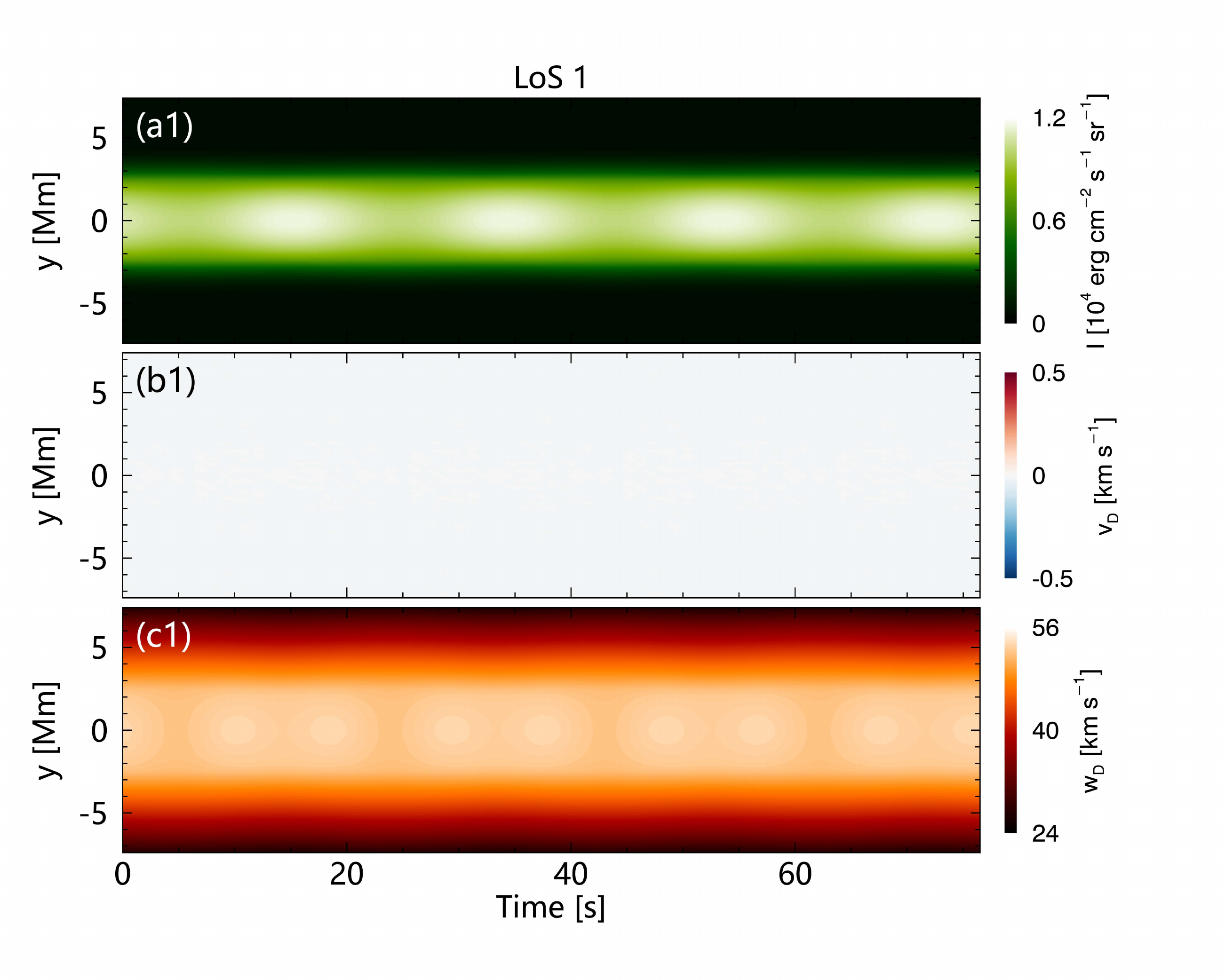}
 \includegraphics[width=0.8\linewidth]{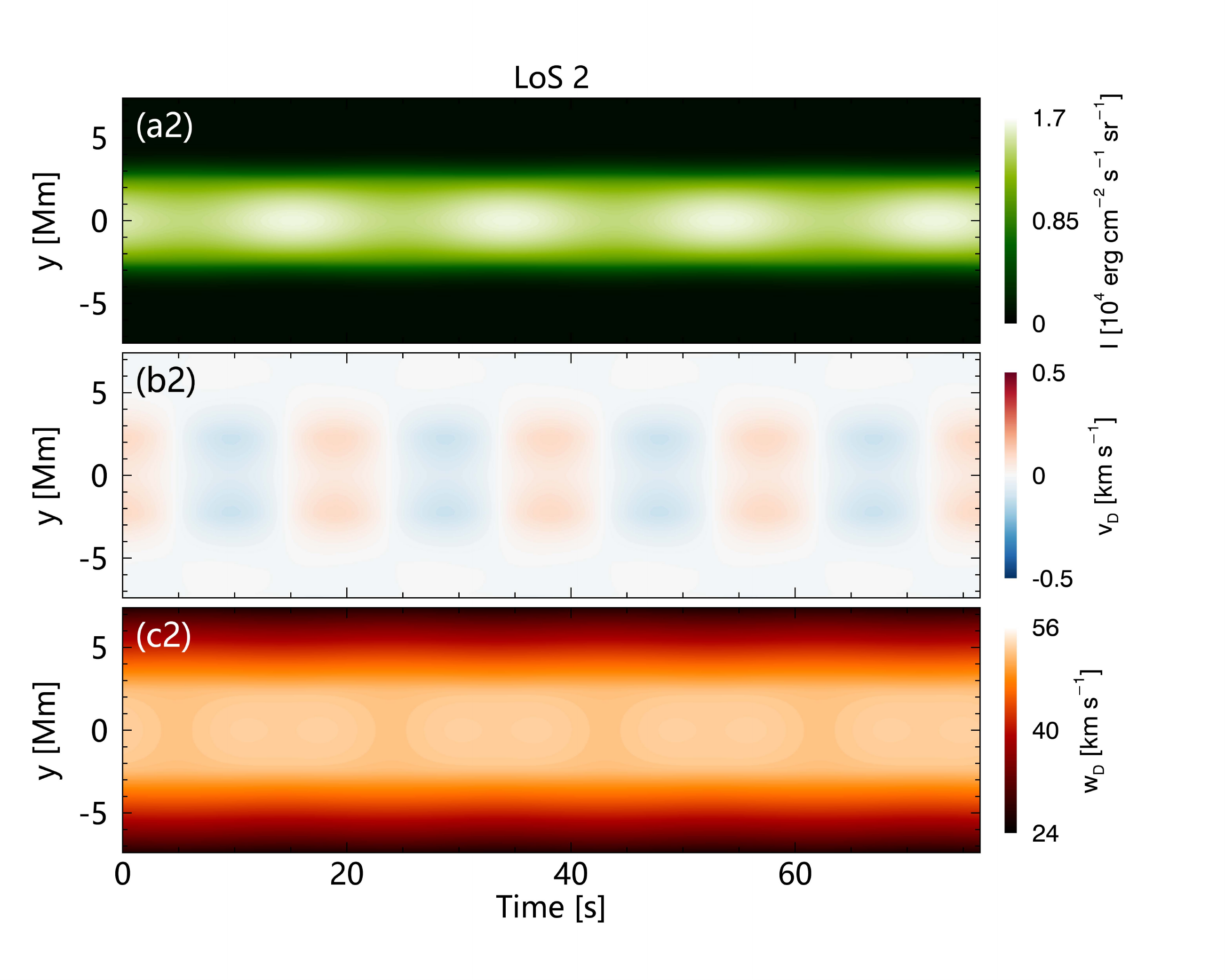}
 \caption{
 Temporal variations of
     (a) intensity $I$,
     (b) Doppler velocity $v_D$,
     and (c) Doppler width $w_D$
 along the slits pertinent to LoS 1 (the upper three panels) and LoS 2 (lower).
 }
 \label{f2}
\end{centering}
\end{figure}

Now we present our forward modeling results 
	and discuss their possible applications to 
	the identification of FSMs using IRIS-like instruments.
Figure \ref{f2} shows the temporal variations of 
	(a) the intensity $I$,
	(b) Doppler velocity $v_D$,
	and (c) Doppler width $w_D$
	along the slits pertinent to
	LoS 1 (the upper three panels) and LoS 2 (lower), respectively.
For both slits, four blob-like intensity enhancements 
    are seen, corresponding to the intervals of strong compression
    (see Figures~\ref{f2}a1 and \ref{f2}a2).
An inspection of the panels labeled (c) shows that the Doppler width variations
    are also similar for both slits.
However, different from the intensity variations, 
    a periodicity of half the wave period ($P/2$) can also be discerned,
    as indicated by the peanut-shaped features. 
The ``peanut kernels'', namely the enhancements in Doppler width,
    are concentrated around the center of a given slit (i.e., $y=0$).
Furthermore, any unshelled peanut sits astride an intensity enhancement. 
The most prominent difference in the spectral signatures between the two slits
    lies in the Doppler shift, which is identically zero in Figure~\ref{f2}b1
    but not in Figure~\ref{f2}b2.
In the former, any pertinent LoS samples the fluid parcels that 
    flow toward and away from the observer in a symmetric manner,
    thereby causing no net Doppler shift.
On the contrary, the contributions from the oppositely moving parcels
    do not exactly cancel out for the slit pertinent to LoS 2
    (see Figure~\ref{f1}a).
One also sees that the local enhancements in the magnitudes of $v_D$ are away from the slit center.
This is understandable because the radial flow speed ($v_r$) is largely distributed this way
    and $v_r$ tends to dominate the velocity projected onto an LoS. 
    
\begin{figure}
  \begin{centering}
  \includegraphics[width=0.8\linewidth]{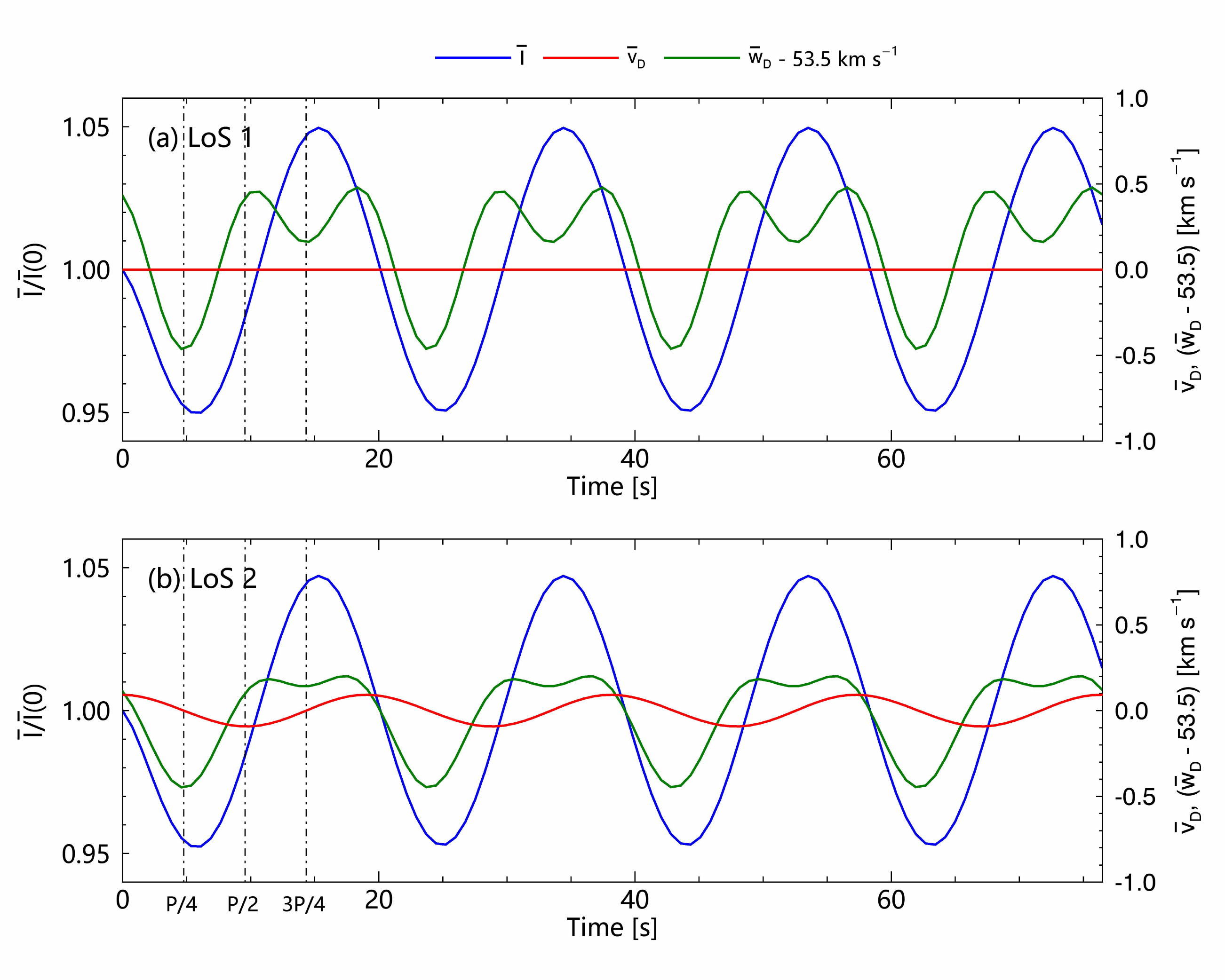}
   \caption{Temporal variations of slit-averaged intensity $\bar{I}$, 
   		Doppler velocity $\bar{v}_D$, 
   		and Doppler width $\bar{w}_D$ (offset by 53.5 km s$^{-1}$).
   By ``slit-averaged'', we mean that these parameters are derived from the spectral profiles averaged over a slit.
   Two slits are examined, one pertinent to LoS 1 (panel a) and the other to LoS 2 (panel b).
   The three vertical dash-dotted lines mark the instants of time when 
       $t=P/4$, $P/2$, and $3P/4$.}
 \label{f3}
\end{centering}
\end{figure}

When real measurements are analyzed, it is customary to average the      
    spectral profiles at individual pixels along a slit to enhance the signal-to-noise ratio.
While the forward-modeling procedure involves little noise, we nonetheless adopt this practice
    by first averaging the spectral profiles in Figure~\ref{f2} at a given time 
    and then applying a Gaussian fitting afterwards.
The time-series of the resulting intensity ($\bar{I}$), Doppler velocity ($\bar{v}_D$)
    and Doppler width ($\bar{w}_D$) are then presented in
    Figure~\ref{f3}, where we examine both (a) the slit pertinent to LoS 1
    and (b) the one pertinent to LoS 2.
For the ease of discussion, the instants of time $t=P/4$, $P/2$, and $3P/4$
    are marked by the vertical dash-dotted lines.    
Note that the examined fundamental mode corresponds to
    the strongest rarefaction (compression)
    at $t=P/4$ ($t=3P/4$), but leads to no compression at $t=0$ or $P/2$.
Note further that the contribution function $G_{\lambda0}$ 
    increases monotonically with temperature in the temperature range we examine.    
Examining the intensities first (the blue curves),
    one sees that they largely follow the density variation ($\Delta N$)
    but nonetheless lag behind $\Delta N$ by $\sim 18^\circ$.
This is somewhat surprising, because one expects a phase difference of $0^\circ$
    given the $T$-dependence of $G_{\lambda0}$ and the fact that $\Delta N$ varies 
    in phase with the temperature variation ($\Delta T$). 
It turns out that this slight phase difference results from the non-equilibrium ionization (NEI) effect,
    namely the ionic fraction of Fe \MyRoman{21} ($f_{\rm XXI}$) cannot respond instantaneously 
    to the temperature variation.
A density-dependence, albeit weak, then results for $f_{\rm XXI}$ and hence for $G_{\lambda 0}$
    (see Equations~\ref{eq_ionic_frac} and \ref{eq_def_G}).
This happens despite that the wave period is rather long and an electron density as high as $5\times 10^{10}~{\rm cm}^{-3}$     
    is adopted.
Nonetheless, the NEI effect is substantially less strong than for     
    the response of the Fe \MyRoman{12} 193 \AA\ line to
    FSMs with much shorter periods in much less dense active region loops (see paper I).
In paper I, we also showed that NEI plays no role in determining the temporal dependence of     
    Doppler widths 
    or Doppler shifts for both Fe \MyRoman{9} 171 \AA\ and Fe \MyRoman{12} 193 \AA.
This is also the case for the Fe \MyRoman{21} 1354 \AA\ line examined here.
As shown by the red curve in Figure~\ref{f3}b, the Doppler velocity
    lags behind the density variation by exactly $90^\circ$,
    and hence behind the intensity variation by $\sim 72^\circ$.
    
\begin{figure}
  \begin{centering}
  \includegraphics[width=\linewidth]{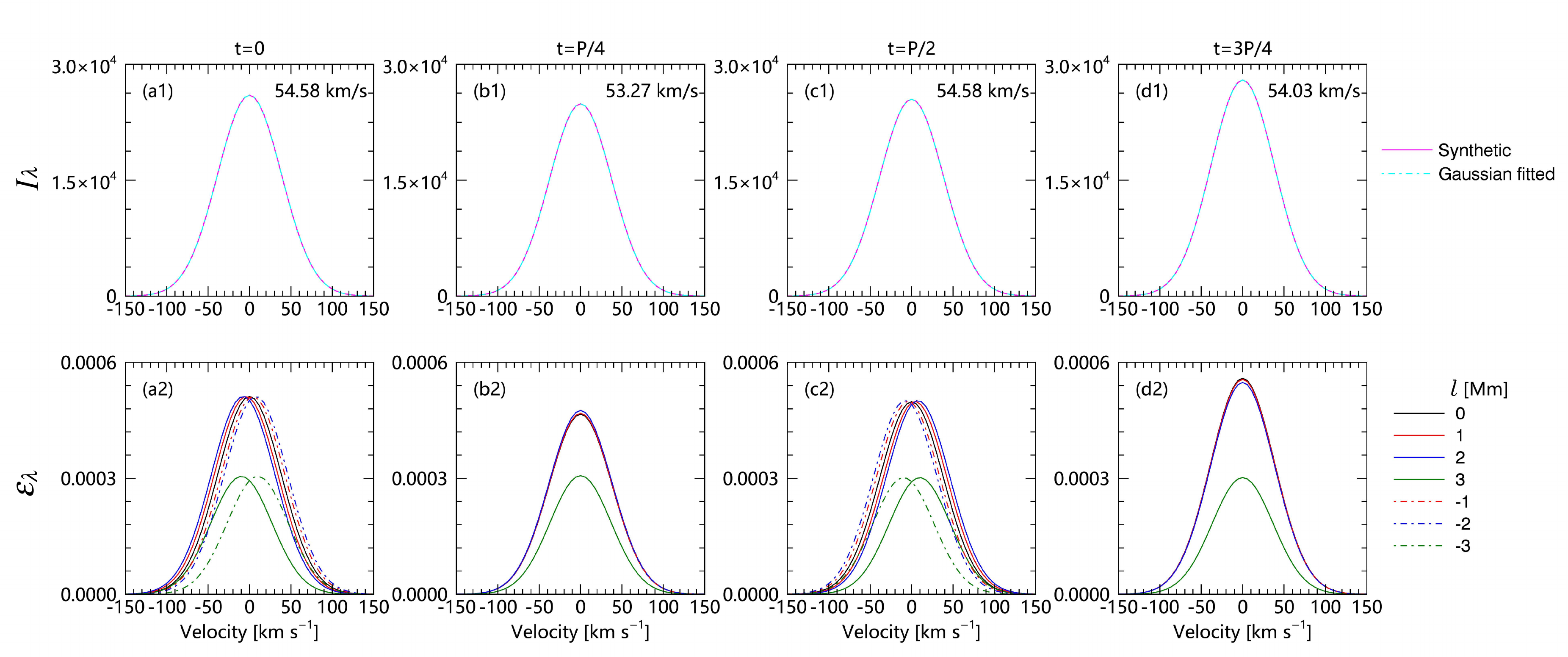}
   \caption{
   Wavelength dependence 
     of the monochromatic intensities $I_\lambda$ (in erg$\rm~cm^{-2}\rm~s^{-1}\rm~\AA^{-1}\rm~sr^{-1}$, the upper row) 
     and that of the monochromatic emissivities $\epsilon_{\lambda}$ (in erg$\rm~cm^{-3}\rm~s^{-1}\rm~\AA^{-1}$, lower) 
     for LoS 1.
   Here the wavelength is presented in velocity units.
   In the upper row, the red curves present the synthetized profiles while the green curves are for the pertinent Gaussian-fitting. 
   The derived Doppler widths are also printed.
   In the lower row, the profiles of $\epsilon_{\lambda}$ at a number of representative locations are presented as labeled.
   Here $l$ denotes the coordinate along the LoS, with $l=0$ corresponding to where the LoS intersects the $y-z$ plane 
       in Figure~\ref{f1}.
   Different columns pertain to different instants of time.}
 \label{f4}
\end{centering}
\end{figure}

The behavior of the Doppler widths, however, is somehow more complicated.
While their local minima appear at both $t=P/4$ and $3P/4$,
    the values at $t=3P/4$ are consistently larger than at $P/4$ for both slits.
This leads to the ``molar''-like features between $t=2/P$ and $P$ in the green curves.
Repeating once every period, these features naturally account for the periodic
    appearance of the ``peanuts'' in Figures~\ref{f2}c1 and \ref{f2}c2.
Then why does the Doppler width behave this way?
This turns out to be a result of the competition between the contribution to the Doppler broadening
    from temperature variations
    and that from the bulk flow.
To explain this further, we note that the temporal variation of the slit-averaged Doppler width
    actually follows closely the behavior of the Doppler width 
    as sampled by the LoS that passes through the slit center.
Therefore Figure~\ref{f4} examines the spectral profiles at four representative
    instants of time (the red curves in the upper row).
Here LoS 1 is taken as an example, because much is the same for LoS 2.    
Furthermore, the green curves pertain to the Gaussian-fitting, and the values
    of the derived Doppler widths are also printed. 
Given that the monochromatic intensity ($I_\lambda$) is an LoS integration of the monochromatic
    emissivity ($\epsilon_\lambda$), we also plot the wavelength-dependence of $\epsilon_\lambda$
    at several representative locations along LoS 1 (the lower row). 
Evidently, the contribution to Doppler broadening from the bulk flow results from 
    the summation of the $\epsilon_\lambda$ profiles shifted in opposite directions
    from the rest wavelength. 
Comparing the left two columns in Figure~\ref{f4}, one sees that in the first quarter of wave period,
    the Doppler broadening weakens and this is because both the temperature
    and flow speeds decrease.
When time proceeds from $P/4$ towards $2/P$, both temperature
    and the flow speeds increase, leading the Doppler width to return to
    its value at $t=0$.
However, when time further proceeds, the electron temperature increases but 
    the flow speeds weaken.
Consequently, the bulk flow tends to reduce the Doppler width between $t=P/2$ and $3P/4$,
     thereby offsetting the tendency for the Doppler width to increase with increasing temperature.
What results is a local minimum in the Doppler width at $3P/4$ that exceeds 
     the corresponding value at $P/4$.

Regarding the Doppler width ($w_D$), 
     two situations are expected.
If the temperature effects far exceed the effects due to the bulk flow, 
     then one expects that $w_D$ will follow a simple sinusoidal shape in phase with 
     the temperature.
If, on the contrary, the bulk flow effects dominate, then $w_D$ is expected to 
     possess a dominant periodicity of $P/2$ by attaining almost identical values
     at $t=P/4$ and $3P/4$.
Our results for the Fe \MyRoman{21} 1354 \AA\ line lie in between these two extremes,
     with the temperature effects being more important given the high temperature pertinent
     to flare loops.
In the case of the Fe \MyRoman{9} 171 \AA\ and \MyRoman{12} 193 \AA\ lines, 
     both \citet{2013A&A...555A..74A} and paper I indicated that the dominant periodicity
     in the Doppler widths is indeed $P/2$ when FSMs in active region loops are examined. 
And this periodicity of $P/2$ takes place because of the dominant role that the bulk flow plays in 
     broadening (or narrowing) the relevant lines. 
     
\begin{figure}
	\begin{centering}
		\includegraphics[width=0.8\linewidth]{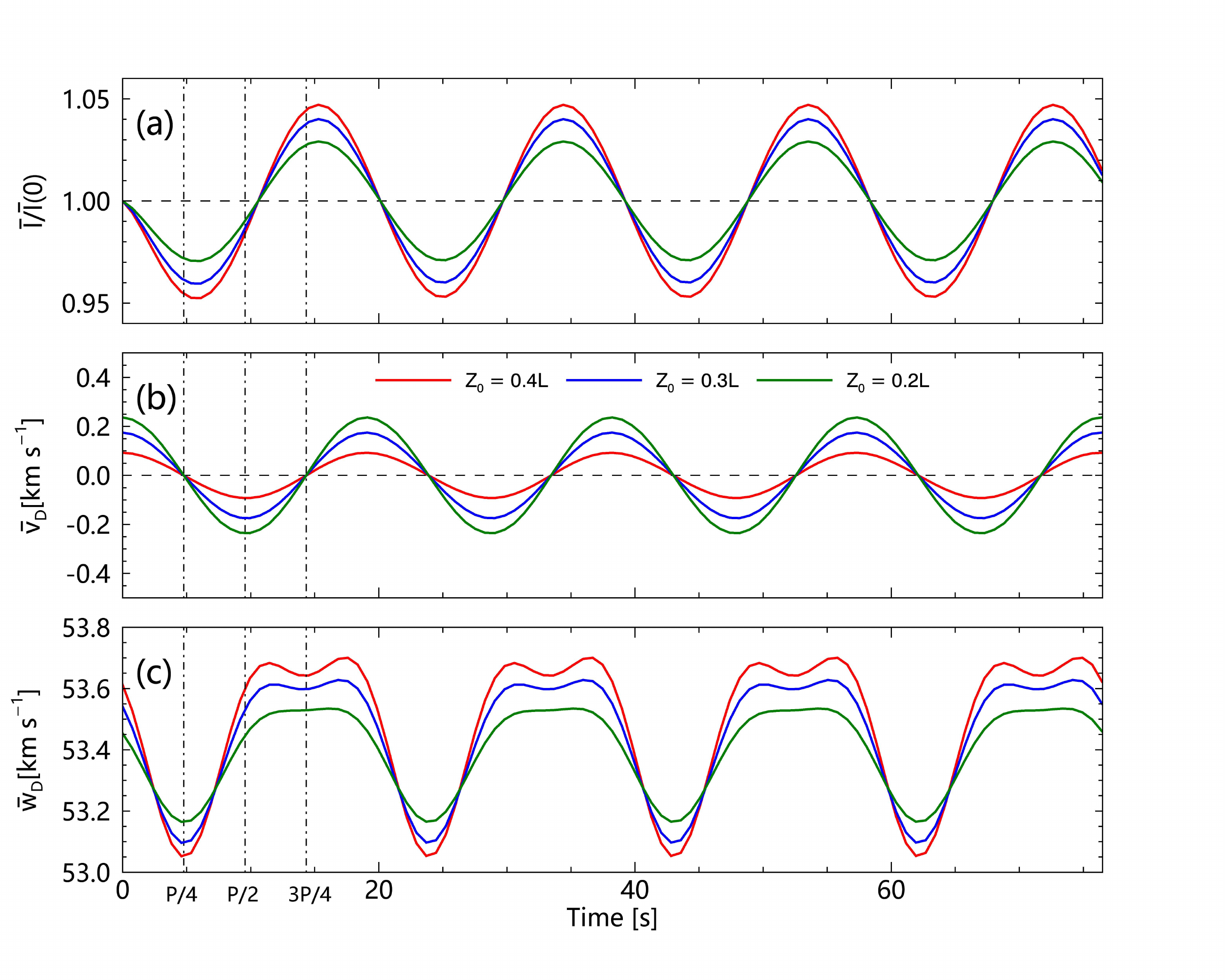}
		
		\caption{Temporal variations of 
		(a) the intensity $\bar{I}$,
		(b) Doppler velocity $\bar{v}_D$,
		and (c) Doppler width $\bar{w}_D$ 
		for slits pertaining to lines of sight that make the same angle ($45^\circ$)
		    with the cylinder as LoS 2 but interest the cylinder axis at different positions ($z_0$).
		The line parameters are derived from the slit-averaged profiles. 
		The three vertical dash-dotted lines correspond to $t=P/4$, $P/2$, and $3P/4$, respectively.
		}
		\label{f5}
	\end{centering}
\end{figure}

\begin{figure}
	\begin{centering}
		\includegraphics[width=0.8\linewidth]{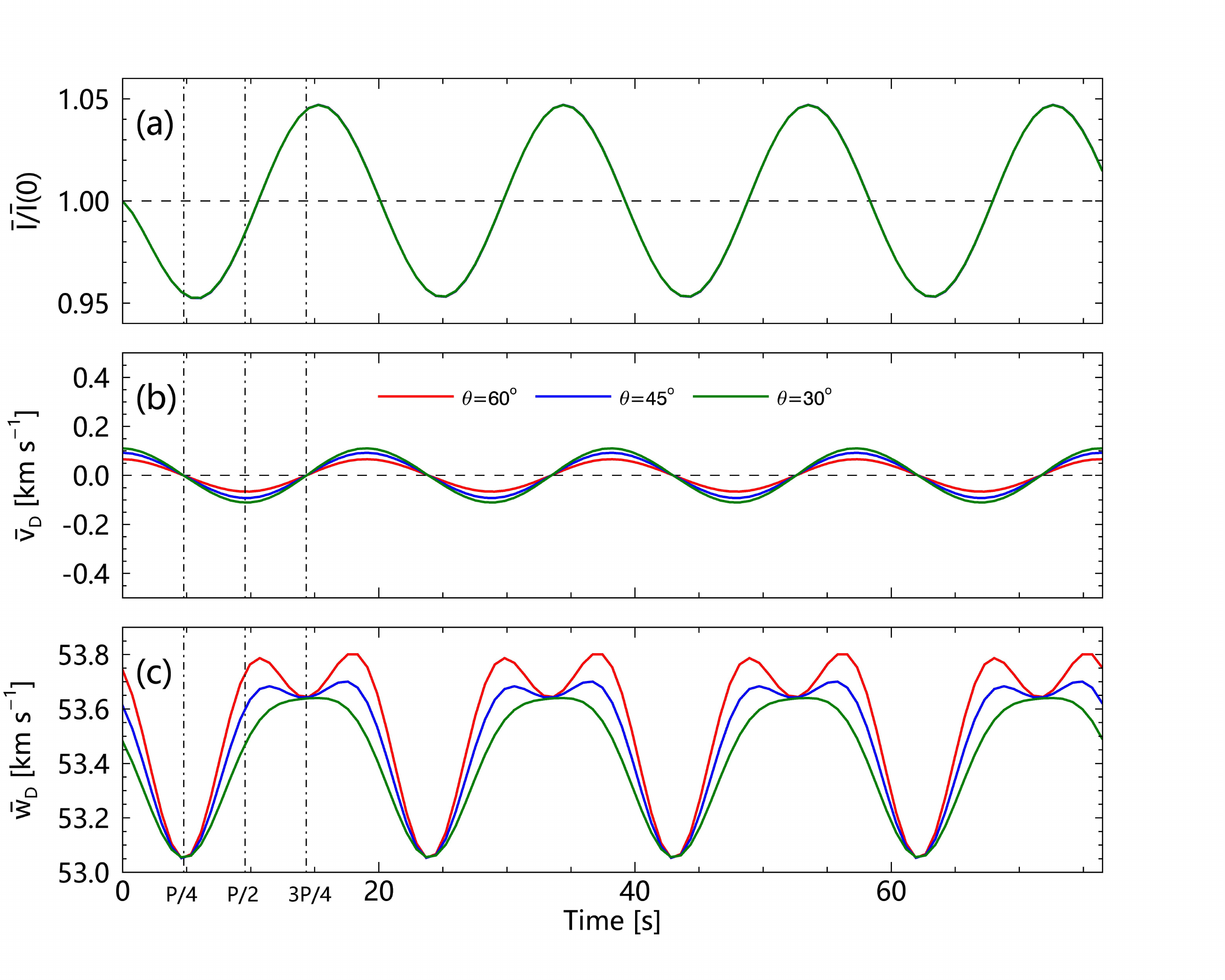}
		
		\caption{
		Similar to Figure~\ref{f5} but for slits pertaining to lines of sight that make 
		    different angles ($\theta$) with the cylinder axis as labeled.
		All slits intersect the $x-z$ plane at $z_0 = 0.4~L$ with $L$ being the cylinder length. 
		}
		\label{f6}
	\end{centering}
\end{figure}

It is also informative to study what happens for slits that pertain to different 
   lines of sight.
This is done in Figures \ref{f5} and \ref{f6} where we examine 
     the temporal variations of 
    (a) the intensity ($\bar{I}$),
    (b) Doppler velocity ($\bar{v}_D$)
    and (c) Doppler width ($\bar{w}_D$).
The slits in Figure~\ref{f5} pertain to lines of sight that are parallel to LoS 2
    but intersect the cylinder axis at different locations ($z_0$) as indicated. 
On the other hand, the slits in Figure~\ref{f6} are relevant for lines of sight that  
    intersect the cylinder axis at the same location but make different angles with the cylinder.
In both figures, the vertical dash-dotted lines mark the instants of time $t=P/4$, $P/2$, and $3P/4$.     
All the spectral parameters are derived from the slit-averaged profiles.
Examine Figure~\ref{f5} first.
One sees that the modulations in both $\bar{I}$ and $\bar{w}_D$
    become increasingly weak with decreasing $z_0$,
    which is understandable because the density and temperature perturbations
    weaken from the loop apex to footpoint.
Somehow the modulation in $\bar{v}_D$ becomes stronger when $z_0$ decreases.
This tendency is understandable because when $z_0$ decreases, 
    the fluid parameters sampled by pertinent lines of sight
    become increasingly asymmetric about the $y-z$ plane.
Furthermore, the phase difference between $\bar{v}_D$ and $\bar{I}$ 
    are all close (although not exactly equal) to $90^\circ$, 
    indicating once again the marginal role of NEI.
More interesting is that, in addition to reducing the magnitude of
    the $\bar{w}_D$ variations, moving away from the loop apex 
    also makes the dip at $3P/4$ less clear
    (Figure~\ref{f5}c).
This is readily understandable with the same reasoning for explaining the molar-shaped feature
    for the red curve here, which is actually reproduced from Figure~\ref{f3}b. 
When time proceeds from $P/2$ to $3P/4$, while the bulk flow effects tend to narrow the line profiles,
    this contribution becomes increasingly weak with decreasing $z_0$.
For $z_0 = 0.2 L$, this narrowing contribution is too weak to compete with the temperature enhancements that
    broaden the profiles, resulting in 
    the plateau around $t = 3P/4$.

Now move on to Figure~\ref{f6}.
One sees from Figure~\ref{f6}a that the intensity variations ($\bar{I}(t)/\bar{I}(t=0)$)
    do not depend on the viewing angle ($\theta$), despite that the absolute values of $\bar{I}(t=0)$ are different
    (not shown).
Given that the Doppler shift $\bar{v}_D$ consistently lags behind the density perturbation by $90^\circ$,
    this independence on viewing angles leads to a phase difference between 
    $\bar{I}$ and $\bar{v}_D$ being consistently $72^\circ$.
(Note that the slit pertinent to $\theta = 45^\circ$
    has been examined in Figure~\ref{f3}b).    
Figure~\ref{f6}b further indicates that the modulations in $\bar{v}_D$ strengthen with decreasing $\theta$.
This can also be explained with Figure~\ref{f2}a: with decreasing $\theta$, the fluid parameters along the pertinent lines of sight
    become increasingly asymmetric about the $y-z$ plane. 
At the expense of this slight enhancement in the Doppler shift modulations, the increasingly strong asymmetry 
    decreases the modulations in the Doppler width  as shown in Figure~\ref{f6}c.
Now the dips at $t=3P/4$ become almost independent of $\theta$ because the flow
    velocities are zero at this instant of time.

From Figures \ref{f5}c and \ref{f6}c one sees a systematic dependence of the Doppler width variations 
    on the viewing angle
    and the axial position where a pertinent LoS intersects the loop.
The systematic variations of the molar-like portions are of particular interest, because 
    they result from the systematic change in the importance 
    of the broadening due to bulk flow relative to that resulting from temperature variations.
We have shown that this should be a rather generic result for such hot emissions lines as Fe {\rm XXI} 1354 \AA\
    provided that thermal effects dominate but not by far dominate the effects due to bulk flow.
In reality, however, the Doppler width variations can hardly behave in the ideal way as depicted in Figures \ref{f5}c
    and \ref{f6}c.
Noise and instrumental resolutions will be an issue, not to mention the difficult task of trend-removal in the time series.
However, the systematic variations of the molar-like portions should also correspond to
    a systematic change in the importance of
    the periodicity of $P/2$ relative to that of the wave period $P$.
Looking for the response of this additional periodicity to such factors as varying viewing angles should be more practical.
For this purpose we present in Figure~\ref{f7} the periodograms of the Doppler width curves
    presented in (a) Figure~\ref{f5}c and (b) Figure~\ref{f6}c.
Note that the FFT power is normalized by its maximum for the ease of comparison between different curves.    
Indeed, one sees that for lines of sight with a fixed direction, the relative importance 
    of the $P/2$ periodicity decreases when an LoS moves away from the loop apex (Figure~\ref{f7}a).
Likewise, for lines of sight that all intersect the loop axis at $z_0 = 0.4 L$,
    this relative importance decreases when an LoS becomes more aligned with the loop.   

{ 
All the results presented so far have been obtained with the ionic fractions by solving
    Equation~\eqref{eq_ionic_frac}.
From the practical viewpoint, one may question whether addressing
   non-Equilibrium ionization (NEI)	is necessary in the first place.
Before answering this question, let us note that 
    one important assumption behind the adopted ionization and recombination rate coefficients (see Equation~\ref{eq_ionic_frac})
    is that electrons are Maxwellian-distributed.
However, a substantial fraction of the electrons in the flare loops examined in T16 is expected to be
   non-thermal as a result of the flaring processes~\citep[see e.g.,][]{2011SSRv..159...19F}. 
While it is possible to address non-Maxwellian-distributed electrons in ionization computations 
   \citep[e.g.,][and references therein]{2018A&A...618A.176D},
   we nonetheless refrain from doing so for the time being.
With this caveat in mind, we repeat all the emission computations by assuming EI from the outset.
In what follows, we first present in Figure~\ref{f8} the contribution function $G_{\lambda 0}$
   (Equation~\ref{eq_def_G}) as a function of the electron temperature ($T$) and density ($N$)
   under the assumption of EI.
Figure~\ref{f8}a examines $G_{\lambda 0}$ for the entire $[T, N]$ range pertaining to both the examined loop itself
   and its ambient.
One sees that $G_{\lambda 0}$ shows essentially no dependence on $N$, and is sharply peaked around 
   $\log_{10} T({\rm K}) \approx 7.1$, meaning that by far the Fe XXI 1354~\AA\ emissions come from the plasmas in
   the core part of the loop provided that Fe XXI is in ionization equilibrium.
The same is true in our computations even though the NEI effects can be discerned.
Figure~\ref{f8}b then presents the portion of the $G_{\lambda 0}$ distribution for the range of electron temperatures
   and densities restricted to the core part of the loop.
In addition, we also plot the evolutionary tracks in the $[N, T, G_{\lambda 0}]$ space of a point initially 
   located at the loop apex ($[r, z] = [0, L/2]$) for both the EI (the red curve) and NEI (black) situations.
This point is representative of the fluid parcels in our computations. 
By construction, the red track starts off from the dot in the middle at $t=0$ and initially moves downward
   before eventually forming a nearly straight line segment on the $G_{\lambda 0}$ surface.
The black track, on the other hand, moves away from the $G_{\lambda 0}$
   surface as the loop system evolves from its initial state, and eventually forms an ellipse.
The deviation of the black track from the red one signifies the NEI effect, and is seen to be rather modest.

The NEI effect is further examined in Figure~\ref{f9}, where the red curves are reproduced from 
   the spectral parameters pertaining to the case with $z_0 = 0.4L$ in Figure~\ref{f5}. 
Comparing them with the blue dashed curves representing the EI case, one sees that, as found in our paper I,
   NEI affects only the line intensity but not the Doppler width or Doppler shift.
Furthermore, Figure~\ref{f9}a indicates that the most discernible difference between
   the NEI and EI curves lies not in the absolute value of the line intensity
   but in the phase difference.
While the first trough in the EI result is located exactly at $t=P/4$, the NEI curve possesses
   a slight delay of $\sim 18^\circ$.    
Repeating the rest of the figures, we reach the same conclusion that    
   NEI plays at most a marginal role for influencing the Fe XXI 1354~\AA\ emissions
   for the adopted equilibrium parameters. 
In fact, a rule of thumb can be readily established by comparing the wave angular frequency ($\omega$) 
   with the relevant 
   ionization ($\nu_{\rm C} = N C_{21}$) and recombination ($\nu_{\rm R} = N R_{21}$) frequencies.
Equation~\eqref{eq_ionic_frac} indicates that the NEI effect becomes negligible when 
   $\omega$ is much smaller than $\nu_{\rm C}$ and $\nu_{\rm R}$.
Note that this requirement is more stringent than that the wave period $P=2\pi/\omega$ be 
   much longer than the ionization and recombination timescales, 
   $\tau_{\rm C}=1/\nu_{\rm C}$ and $\tau_{\rm R}=1/\nu_{\rm R}$.
For instance,       
   the equilibrium parameters at the loop apex lead to
   $\tau_{\rm C} \approx 1.6$~sec and $\tau_{\rm R} \approx 0.8$~sec, which are indeed much shorter
   than the period $P=19.1$~sec. 
The angular frequency $\omega$, on the other hand, is not that small when compared with $\nu_{\rm C}$
   and $\nu_{\rm R}$, leading to the modest but discernible effect that NEI plays
   in determining the ionic fractions.   
}
    
\section{Summary and Concluding Remarks}
\label{summary}
This work was inspired by the recent IRIS detection of spectral signatures in 
    the Fe \MyRoman{21} 1354~\AA\ line of a fundamental standing fast sausage mode (FSM) trapped
    in flare loops~\citep[][T16]{2016ApJ...823L..16T}.
Starting with the solutions for linear FSMs in a straight cylinder with equilibrium parameters
    largely compatible with T16, 
    we have forward-modeled the modulations to the spectral parameters
    of this flare line.
Our results suggest     
    the following signatures in Fe \MyRoman{21} 1354~\AA\ measurements
    for fundamental FSMs.
One. The intensity variations consistently possess the same period as the FSM.
Two. When the loop is sampled at non-$90^\circ$ angles with respect to its axis,
    the variations in the Doppler shift can be seen and have a nearly $90^\circ$ phase difference
    with respect to the intensity modulations.
Three. In general the variations in the Doppler width possess an irregular shape due to their asymmetry
    in the first and second halves of a wave period ($P$).
    This asymmetry results in a secondary periodicity at $P/2$, whose importance relative to 
    the dominant periodicity of $P$ shows a systematic dependence on the 
    viewing angle and the axial location where a slit intersects the loop.

Let us remark that an additional signature is that 
    $P$ should be as short as being comparable to the transverse fast time ($R/v_{\rm fi}$), 
    by which we mean the time
    it takes perpendicularly propagating fast waves to traverse the loop in the transverse direction
    \citep[e.g.][]{2016ApJ...833..114C}.
This fourth signature was not mentioned because it has been well-established in the literature
    \citep[see e.g.,][]{1978SoPh...58..165M, 1983SoPh...88..179E}.
Note that $v_{\rm fi}$ should be evaluated with the equilibrium parameters in the loop.    
Note further that this property of trapped fundamental FSMs is not to be confused with the notion that their periods
    are no longer than twice the longitudinal Alfv\'en time ($2L/v_{\rm Ae}$), which actually refers to
    the time it takes the Alfv\'en waves in the ambient corona to traverse the length
    of the loop.
The axial phase speed of trapped FSMs equals the ambient Alfv\'en speed at the cutoff axial wavenumber ($k$),
    below which FSMs become leaky.
Given that the frequency of a trapped FSM increases with increasing $k$ and that $k=\pi/L$ for the axial fundamental mode,
    one finds that indeed $P$ is no longer than $2L/v_{\rm Ae}$.
However, theories indicate that $2L/v_{\rm Ae}$ is on the same order as $R/v_{\rm fi}$, 
    and evaluates to $\sim 2.6 R/v_{\rm fi}$ for loops with transversely discontinuous profiles~\citep[e.g.,][]{2007AstL...33..706K}.
While in reality the equilibrium parameters of coronal loops are unlikely to be distributed in a discontinuous manner,
    taking continuous profiles into account still leads to the same conclusion~\citep{2016ApJ...833..114C}.
In fact, rather than directly using Signature 4, T16 estimated the axial phase speed and 
    employed the derived large value as one piece of evidence that 
    the oscillations they measured with IRIS are likely to be caused by a fundamental FSM.
The considerable difference between the loop length $L$ and radius $R$, by roughly an order-of-magnitude even for the flare loops that T16 examined,
    means that loops need to be much denser than the ambient for FSMs to be trapped~\citep[e.g.,][]{2004ApJ...600..458A}.
This was found not to be an issue for the dense flare loop examined in T16, though, as indicated by their DEM-based density estimation.

Signatures 1 and 2 are also consistent with the IRIS measurements.
Indeed, T16 found a roughly periodic intensity variation, which has a phase difference of $\sim 90^\circ$ relative to
     the Doppler shift variations.
We note that the phase difference we derived is $\sim 72^\circ$, 
     which is slightly different from $90^\circ$ 
     because Fe \MyRoman{21} is not in perfect ionization equilibrium despite
     we adopted the measured loop density ($\sim 5\times 10^{10}~{\rm cm}^{-3}$).
The difference between the computed and measured values of the phase difference 
     may come from the fact that the measured period ($\sim 25~{\rm sec}$) is slightly longer
     than computed here ($\sim 19.1~{\rm sec}$).
Alternatively, it may be that the phase difference cannot be measured with precision because 
     the cadence ($\sim 5.2~{\rm sec}$) is a substantial fraction of the measured period.

\begin{figure}
	\begin{centering}
		\includegraphics[width=0.8\linewidth]{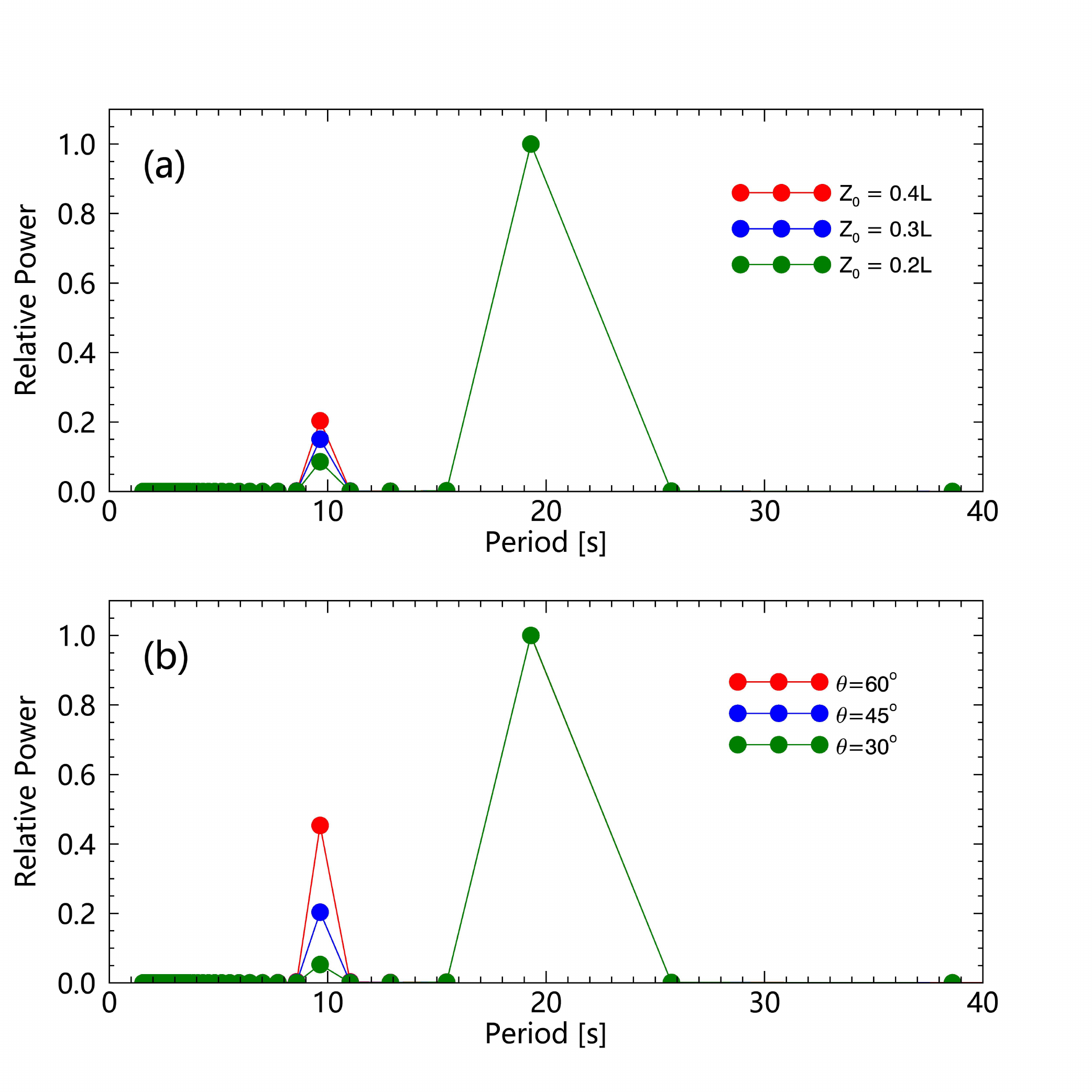}
		\caption{Fourier spectra for the Doppler width variations examined 
		in (a) Figure \ref{f5} and (b) Figure \ref{f6}. 
		Any spectrum is normalized by its maximum, and hence the label ``relative power''.
		}
		\label{f7}
	\end{centering}
\end{figure}

\begin{figure}
	\begin{centering}
		\includegraphics[width=0.65\linewidth]{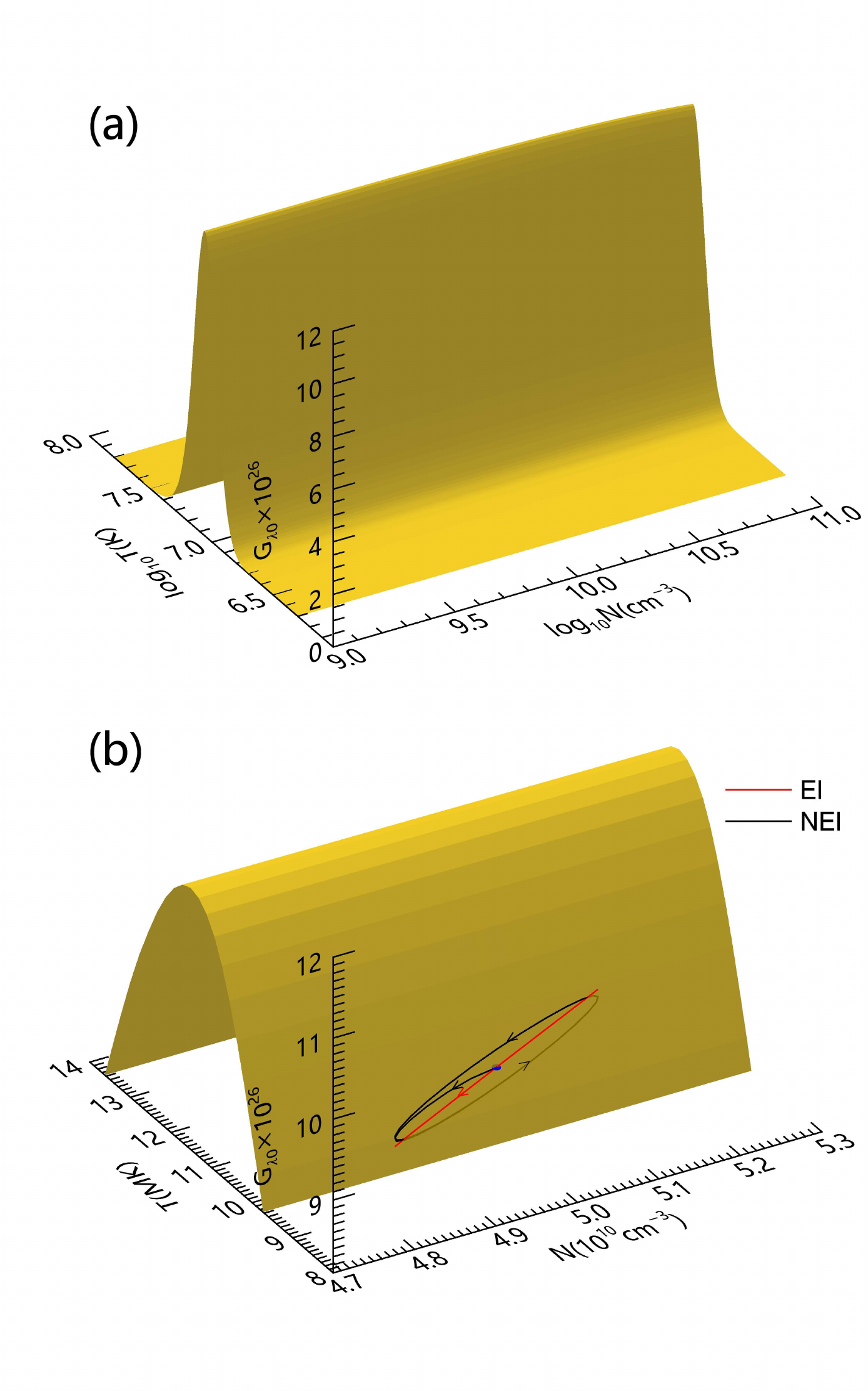}
		\caption{Contribution function ($G_{\lambda 0}$) under the assumption of equilibrium ionization (EI)
			for (a) the entire range of the electron density ($N$) and temperature ($T$) examined in this study,
			and (b) a limited range pertaining to the core of the loop. 
			The values of $G_{\lambda 0}$ (in units of ${\rm erg}~{\rm cm}^{3}~{\rm s}^{-1}$) are returned from 
			   the function \textbf{g\_of\_t} in the Chianti package. 
			The curves in panel b represent the trajectories in this three-dimensional space 
			   of a point initially located at the loop apex
			   when EI (the red curve)
			   and NEI (black) are assumed. 
			While they both start off from the dot in the middle at $t=0$, the red track 
			   initially moves downward and forms a nearly straight line segment on the $G_{\lambda 0}$ surface, 
			   whereas the black track deviates from the $G_{\lambda 0}$ surface and eventually forms
			   an ellipse.
			The difference between the two evolutionary tracks signifies the NEI effect.    
		}
		\label{f8}
	\end{centering}
\end{figure}

\begin{figure}
	\begin{centering}
		\includegraphics[width=0.7\linewidth]{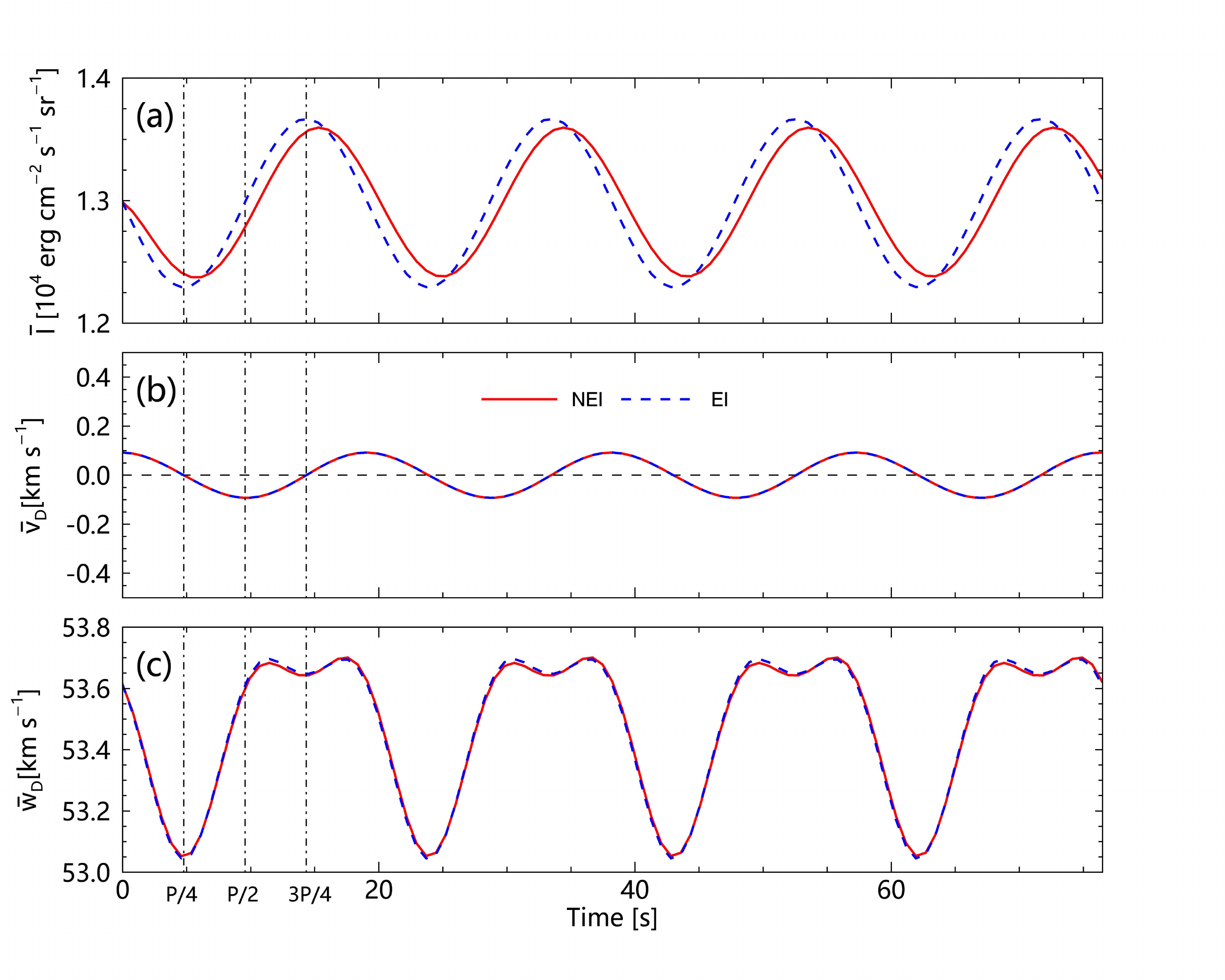}
		\caption{Similar to Figure~\ref{f5} but concerning only the slit with $z_0 = 0.4L$. 
		The red solid (blue dashed) curves pertain to the case where
		    the ionic fractions of Fe XXI are found with the full Equation~\eqref{eq_ionic_frac}
		    (by assuming ionization equilibrium). 
		}
		\label{f9}
	\end{centering}
\end{figure}

The cadence issue makes the comparison of Signature 3 with the IRIS measurements impossible.
As shown in our Figure~\ref{f7}, the cadence of a spectral instrument needs to be much shorter than
     the wave period for the secondary periodicity to be discerned in the Doppler width variations.
If boldly taking $25~{\rm sec}$ as being representative of standing FSMs in flare loops,
     a cadence of $\sim 2~{\rm sec}$ will be necessary.
An additional issue arises if we further compare the forward modeled line parameters with T16.
By choosing a small wave amplitude, 
     we made sure that the intensities vary by only several percent to be compatible with T16.
However, the measured Doppler shift variations are substantially stronger than those in the Doppler widths (Figure~3B in T16),
     whereas the two have roughly the same magnitude in our computations (see e.g., Figure~5). 
Many factors could have contributed to the cause of this discrepancy, and we name but one.
As has been briefly discussed, adopting a continuous equilibrium profile is necessary to correctly compute
     the line parameters, the intensity in particular.
This, however, introduces such free parameters as the width and profile description in the transition layer.
Examining the effects of these parameters on the synthetic emissions is certainly warranted,
     but is nonetheless left for a future study.

\acknowledgments
This work is supported by
    the National Natural Science Foundation of China (41474149, U1831112, 41604145, 41674172, 11761141002).
ZH is supported by the Young Scholar Program of Shandong University Weihai (2017WHWLJH07).    
This work is also supported by the Open Research Program of
   the Key Laboratory of Solar Activity of National Astronomical Observatories of China
   (BL: KLSA201801).
We also acknowledge the International Space Science
   Institute Beijing (ISSI-BJ) for supporting the international team
   ``MHD Seismology of the Solar Corona".   
CHIANTI is a collaborative project involving George Mason University, 
    the University of Michigan (USA) and the University of Cambridge (UK).

\bibliographystyle{apj}
\bibliography{Saus}

\begin{thebibliography}{}
\expandafter\ifx\csname natexlab\endcsname\relax\def\natexlab#1{#1}\fi

\bibitem[{{Antolin} \& {Van Doorsselaere}(2013)}]{2013A&A...555A..74A}
{Antolin}, P., \& {Van Doorsselaere}, T. 2013, \aap, 555, A74

\bibitem[{{Aschwanden} {et~al.}(2004){Aschwanden}, {Nakariakov}, \&
  {Melnikov}}]{2004ApJ...600..458A}
{Aschwanden}, M.~J., {Nakariakov}, V.~M., \& {Melnikov}, V.~F. 2004, \apj, 600,
  458

\bibitem[{{Banerjee} {et~al.}(2007){Banerjee}, {Erd{\'e}lyi}, {Oliver}, \&
  {O'Shea}}]{2007SoPh..246....3B}
{Banerjee}, D., {Erd{\'e}lyi}, R., {Oliver}, R., \& {O'Shea}, E. 2007,
  \solphys, 246, 3

\bibitem[{{Bennett} {et~al.}(1999){Bennett}, {Roberts}, \&
  {Narain}}]{1999SoPh..185...41B}
{Bennett}, K., {Roberts}, B., \& {Narain}, U. 1999, \solphys, 185, 41

\bibitem[{{Cally}(1986)}]{1986SoPh..103..277C}
{Cally}, P.~S. 1986, \solphys, 103, 277

\bibitem[{{Cally} \& {Xiong}(2018)}]{2018JPhA...51b5501C}
{Cally}, P.~S., \& {Xiong}, M. 2018, Journal of Physics A Mathematical General,
  51, 025501

\bibitem[{{Chen} {et~al.}(2015){Chen}, {Li}, {Xiong}, {Yu}, \&
  {Guo}}]{2015ApJ...812...22C}
{Chen}, S.-X., {Li}, B., {Xiong}, M., {Yu}, H., \& {Guo}, M.-Z. 2015, \apj,
  812, 22

\bibitem[{{Chen} {et~al.}(2016){Chen}, {Li}, {Xiong}, {Yu}, \&
  {Guo}}]{2016ApJ...833..114C}
---. 2016, \apj, 833, 114

\bibitem[{{Cooper} {et~al.}(2003){Cooper}, {Nakariakov}, \&
  {Tsiklauri}}]{2003A&A...397..765C}
{Cooper}, F.~C., {Nakariakov}, V.~M., \& {Tsiklauri}, D. 2003, \aap, 397, 765

\bibitem[{{De Moortel} \& {Nakariakov}(2012)}]{2012RSPTA.370.3193D}
{De Moortel}, I., \& {Nakariakov}, V.~M. 2012, Philosophical Transactions of
  the Royal Society of London Series A, 370, 3193

\bibitem[{{Del Zanna} {et~al.}(2015){Del Zanna}, {Dere}, {Young}, {Landi}, \&
  {Mason}}]{2015A&A...582A..56D}
{Del Zanna}, G., {Dere}, K.~P., {Young}, P.~R., {Landi}, E., \& {Mason}, H.~E.
  2015, \aap, 582, A56

\bibitem[{{Dzif{\v c}{\'a}kov{\'a}} \&
  {Karlick{\'y}}(2018)}]{2018A&A...618A.176D}
{Dzif{\v c}{\'a}kov{\'a}}, E., \& {Karlick{\'y}}, M. 2018, \aap, 618, A176

\bibitem[{{Edwin} \& {Roberts}(1983)}]{1983SoPh...88..179E}
{Edwin}, P.~M., \& {Roberts}, B. 1983, \solphys, 88, 179

\bibitem[{{Erd{\'e}lyi} \& {Fedun}(2007)}]{2007SoPh..246..101E}
{Erd{\'e}lyi}, R., \& {Fedun}, V. 2007, \solphys, 246, 101

\bibitem[{{Fletcher} {et~al.}(2011){Fletcher}, {Dennis}, {Hudson}, {Krucker},
  {Phillips}, {Veronig}, {Battaglia}, {Bone}, {Caspi}, {Chen}, {Gallagher},
  {Grigis}, {Ji}, {Liu}, {Milligan}, \& {Temmer}}]{2011SSRv..159...19F}
{Fletcher}, L., {Dennis}, B.~R., {Hudson}, H.~S., {et~al.} 2011, \ssr, 159, 19

\bibitem[{{Gruszecki} {et~al.}(2012){Gruszecki}, {Nakariakov}, \& {Van
  Doorsselaere}}]{2012A&A...543A..12G}
{Gruszecki}, M., {Nakariakov}, V.~M., \& {Van Doorsselaere}, T. 2012, \aap,
  543, A12

\bibitem[{{Inglis} {et~al.}(2009){Inglis}, {van Doorsselaere}, {Brady}, \&
  {Nakariakov}}]{2009A&A...503..569I}
{Inglis}, A.~R., {van Doorsselaere}, T., {Brady}, C.~S., \& {Nakariakov}, V.~M.
  2009, \aap, 503, 569

\bibitem[{{Kaneda} {et~al.}(2018){Kaneda}, {Misawa}, {Iwai}, {Masuda},
  {Tsuchiya}, {Katoh}, \& {Obara}}]{2018ApJ...855L..29K}
{Kaneda}, K., {Misawa}, H., {Iwai}, K., {et~al.} 2018, \apjl, 855, L29

\bibitem[{{Khongorova} {et~al.}(2012){Khongorova}, {Mikhalyaev}, \&
  {Ruderman}}]{2012SoPh..280..153K}
{Khongorova}, O.~V., {Mikhalyaev}, B.~B., \& {Ruderman}, M.~S. 2012, \solphys,
  280, 153

\bibitem[{{Ko} {et~al.}(2010){Ko}, {Raymond}, {Vr{\v s}nak}, \&
  {Vuji{\'c}}}]{2010ApJ...722..625K}
{Ko}, Y.-K., {Raymond}, J.~C., {Vr{\v s}nak}, B., \& {Vuji{\'c}}, E. 2010,
  \apj, 722, 625

\bibitem[{{Kolotkov} {et~al.}(2018){Kolotkov}, {Nakariakov}, \&
  {Kontar}}]{2018ApJ...861...33K}
{Kolotkov}, D.~Y., {Nakariakov}, V.~M., \& {Kontar}, E.~P. 2018, \apj, 861, 33

\bibitem[{{Kolotkov} {et~al.}(2015){Kolotkov}, {Nakariakov}, {Kupriyanova},
  {Ratcliffe}, \& {Shibasaki}}]{2015A&A...574A..53K}
{Kolotkov}, D.~Y., {Nakariakov}, V.~M., {Kupriyanova}, E.~G., {Ratcliffe}, H.,
  \& {Shibasaki}, K. 2015, \aap, 574, A53

\bibitem[{{Kopylova} {et~al.}(2007){Kopylova}, {Melnikov}, {Stepanov}, {Tsap},
  \& {Goldvarg}}]{2007AstL...33..706K}
{Kopylova}, Y.~G., {Melnikov}, A.~V., {Stepanov}, A.~V., {Tsap}, Y.~T., \&
  {Goldvarg}, T.~B. 2007, Astronomy Letters, 33, 706

\bibitem[{{Landi} {et~al.}(2006){Landi}, {Del Zanna}, {Young}, {Dere}, {Mason},
  \& {Landini}}]{2006ApJS..162..261L}
{Landi}, E., {Del Zanna}, G., {Young}, P.~R., {et~al.} 2006, \apjs, 162, 261

\bibitem[{{Li} {et~al.}(2014){Li}, {Chen}, {Xia}, \&
  {Yu}}]{2014A&A...568A..31L}
{Li}, B., {Chen}, S.-X., {Xia}, L.-D., \& {Yu}, H. 2014, \aap, 568, A31

\bibitem[{{Li} {et~al.}(2013){Li}, {Habbal}, \& {Chen}}]{2013ApJ...767..169L}
{Li}, B., {Habbal}, S.~R., \& {Chen}, Y. 2013, \apj, 767, 169

\bibitem[{{McLaughlin} {et~al.}(2018){McLaughlin}, {Nakariakov}, {Dominique},
  {Jel{\'{\i}}nek}, \& {Takasao}}]{2018SSRv..214...45M}
{McLaughlin}, J.~A., {Nakariakov}, V.~M., {Dominique}, M., {Jel{\'{\i}}nek},
  P., \& {Takasao}, S. 2018, \ssr, 214, 45

\bibitem[{{Meerson} {et~al.}(1978){Meerson}, {Sasorov}, \&
  {Stepanov}}]{1978SoPh...58..165M}
{Meerson}, B.~I., {Sasorov}, P.~V., \& {Stepanov}, A.~V. 1978, \solphys, 58,
  165

\bibitem[{{Melnikov} {et~al.}(2005){Melnikov}, {Reznikova}, {Shibasaki}, \&
  {Nakariakov}}]{2005A&A...439..727M}
{Melnikov}, V.~F., {Reznikova}, V.~E., {Shibasaki}, K., \& {Nakariakov}, V.~M.
  2005, \aap, 439, 727

\bibitem[{{Nakariakov} {et~al.}(2018){Nakariakov}, {Anfinogentov},
  {Storozhenko}, {Kurochkin}, {Bogod}, {Sharykin}, \&
  {Kaltman}}]{2018ApJ...859..154N}
{Nakariakov}, V.~M., {Anfinogentov}, S., {Storozhenko}, A.~A., {et~al.} 2018,
  \apj, 859, 154

\bibitem[{{Nakariakov} {et~al.}(2012){Nakariakov}, {Hornsey}, \&
  {Melnikov}}]{2012ApJ...761..134N}
{Nakariakov}, V.~M., {Hornsey}, C., \& {Melnikov}, V.~F. 2012, \apj, 761, 134

\bibitem[{{Nakariakov} \& {Melnikov}(2009)}]{2009SSRv..149..119N}
{Nakariakov}, V.~M., \& {Melnikov}, V.~F. 2009, \ssr, 149, 119

\bibitem[{{Nakariakov} {et~al.}(2003){Nakariakov}, {Melnikov}, \&
  {Reznikova}}]{2003A&A...412L...7N}
{Nakariakov}, V.~M., {Melnikov}, V.~F., \& {Reznikova}, V.~E. 2003, \aap, 412,
  L7

\bibitem[{{Nakariakov} \& {Verwichte}(2005)}]{2005LRSP....2....3N}
{Nakariakov}, V.~M., \& {Verwichte}, E. 2005, Living Reviews in Solar Physics,
  2, 3

\bibitem[{{Nakariakov} {et~al.}(2016){Nakariakov}, {Pilipenko}, {Heilig},
  {Jel{\'{\i}}nek}, {Karlick{\'y}}, {Klimushkin}, {Kolotkov}, {Lee},
  {Nistic{\`o}}, {Van Doorsselaere}, {Verth}, \&
  {Zimovets}}]{2016SSRv..200...75N}
{Nakariakov}, V.~M., {Pilipenko}, V., {Heilig}, B., {et~al.} 2016, \ssr, 200,
  75

\bibitem[{{Rosenberg}(1970)}]{1970A&A.....9..159R}
{Rosenberg}, H. 1970, \aap, 9, 159

\bibitem[{{Shen} {et~al.}(2013){Shen}, {Reeves}, {Raymond}, {Murphy}, {Ko},
  {Lin}, {Miki{\'c}}, \& {Linker}}]{2013ApJ...773..110S}
{Shen}, C., {Reeves}, K.~K., {Raymond}, J.~C., {et~al.} 2013, \apj, 773, 110

\bibitem[{{Shi} {et~al.}(2019){Shi}, {Li}, {Van Doorsselaere}, {Chen}, \&
  {Huang}}]{2019ApJ...870...99S}
{Shi}, M., {Li}, B., {Van Doorsselaere}, T., {Chen}, S.-X., \& {Huang}, Z.
  2019, \apj, 870, 99

\bibitem[{{Su} {et~al.}(2012){Su}, {Shen}, {Liu}, {Liu}, \&
  {Mao}}]{2012ApJ...755..113S}
{Su}, J.~T., {Shen}, Y.~D., {Liu}, Y., {Liu}, Y., \& {Mao}, X.~J. 2012, \apj,
  755, 113

\bibitem[{{Terra-Homem} {et~al.}(2003){Terra-Homem}, {Erd{\'e}lyi}, \&
  {Ballai}}]{2003SoPh..217..199T}
{Terra-Homem}, M., {Erd{\'e}lyi}, R., \& {Ballai}, I. 2003, \solphys, 217, 199

\bibitem[{{Terradas} {et~al.}(2005){Terradas}, {Oliver}, \&
  {Ballester}}]{2005A&A...441..371T}
{Terradas}, J., {Oliver}, R., \& {Ballester}, J.~L. 2005, \aap, 441, 371

\bibitem[{{Tian} {et~al.}(2016){Tian}, {Young}, {Reeves}, {Wang}, {Antolin},
  {Chen}, \& {He}}]{2016ApJ...823L..16T}
{Tian}, H., {Young}, P.~R., {Reeves}, K.~K., {et~al.} 2016, \apjl, 823, L16

\bibitem[{{Van Doorsselaere} {et~al.}(2016{\natexlab{a}}){Van Doorsselaere},
  {Antolin}, {Yuan}, {Reznikova}, \& {Magyar}}]{2016FrASS...3....4V}
{Van Doorsselaere}, T., {Antolin}, P., {Yuan}, D., {Reznikova}, V., \&
  {Magyar}, N. 2016{\natexlab{a}}, Frontiers in Astronomy and Space Sciences,
  3, 4

\bibitem[{{Van Doorsselaere} {et~al.}(2016{\natexlab{b}}){Van Doorsselaere},
  {Kupriyanova}, \& {Yuan}}]{2016SoPh..291.3143V}
{Van Doorsselaere}, T., {Kupriyanova}, E.~G., \& {Yuan}, D. 2016{\natexlab{b}},
  \solphys, 291, 3143

\bibitem[{{Yu} {et~al.}(2015){Yu}, {Li}, {Chen}, \&
  {Guo}}]{2015ApJ...814...60Y}
{Yu}, H., {Li}, B., {Chen}, S.-X., \& {Guo}, M.-Z. 2015, \apj, 814, 60

\bibitem[{{Yu} {et~al.}(2013){Yu}, {Nakariakov}, {Selzer}, {Tan}, \&
  {Yan}}]{2013ApJ...777..159Y}
{Yu}, S., {Nakariakov}, V.~M., {Selzer}, L.~A., {Tan}, B., \& {Yan}, Y. 2013,
  \apj, 777, 159

\bibitem[{{Yu} {et~al.}(2016){Yu}, {Nakariakov}, \&
  {Yan}}]{2016ApJ...826...78Y}
{Yu}, S., {Nakariakov}, V.~M., \& {Yan}, Y. 2016, \apj, 826, 78

\bibitem[{{Zajtsev} \& {Stepanov}(1975)}]{1975IGAFS..37....3Z}
{Zajtsev}, V.~V., \& {Stepanov}, A.~V. 1975, Issledovaniia Geomagnetizmu
  Aeronomii i Fizike Solntsa, 37, 3

\end{thebibliography}


\clearpage 

\clearpage 

\clearpage 

\clearpage 

\clearpage 

\clearpage 

\clearpage 

\clearpage 

\clearpage 

\end{document}